\documentclass[aps,prb,twocolumn,showpacs,superscriptaddress, longbibliography]{revtex4-1}

\usepackage{xcolor, graphicx}
\usepackage{amssymb,amsmath, amsfonts, bm}
\usepackage{newtxtext,newtxmath}

\usepackage{dcolumn}
\usepackage[none]{hyphenat} 
\usepackage{tabularx}

\begin{document}

\title{Nodal surface and persistent spin texture in a Weyl semimetal without mirror symmetry}
\author{Wujun Shi}
\affiliation{School of Physical Science and Technology, ShanghaiTech University, Shanghai 200031, China}
\author{Gang Li}
\affiliation{School of Physical Science and Technology, ShanghaiTech University, Shanghai 200031, China}
\affiliation{\mbox{ShanghaiTech Laboratory for Topological Physics, ShanghaiTech University, Shanghai 200031, China}}

\date{\today}
\begin{abstract}
By utilizing symmetry analysis and electronic structure calculations, we investigated the low-temperature orthorhombic phase of Ag$_{2}$Se in ${\cal SG}$~17. In addition to the discovery of a nodal plane at $k_{z}=\pi$ protected by the joint operation of time-reversal (${\cal T}$) and the 2-fold screw rotation $S_{2z}$, we found 24 Weyl points mainly residing at the $k_{y}=0$ plane with notable Fermi arc and large quasiparticle interference pattern (QPI). Due to the absence of mirror symmetry, a pair of Weyl points with opposite chirality reside at different binding energies, which makes this system an excellent material candidate for realizing the novel chiral anomaly related phenomenon, such as the quantized circular photogalvanic and the chiral magnetic effects.
Furthermore, we also reveal the striking spin textures at $k_{z}=\pi$ plane which demonstrates, in a large region of the surface Brillouin Zone, a direction-selective spin polarization, which has a strong implication to spintronic applications.
\end{abstract}

\maketitle

\section{Introduction}
Equipped with the predicting power of the topological band theory, the study of topological materials has revolutionized the conventional viewpoint on the state of matter and triggered a great interest in the nontrivial Berry curvature of the Bloch wavefunction\cite{Wan2011,
Weng2015, Huang2015, Lv2015TaAs,
Xu2015TaAs,
Wang2015MoTe2, Sun2015MoTe2, Soluyanov2015WTe2, jiang2017signature,
Vafek2014,
Wang_Heusler_2016,
armitage2017weyl}.
Among the many new states of matter, the study of three-dimensional (3D) Weyl semimetal (WSM) has emerged as a new research discipline known as topological semimetal,
where the lower-energy excitations around a degenerate Weyl point (WP) between two energy bands are well described as a 3D Weyl fermion.
In solids, WPs always appear in pairs of opposite chiralities with quantized Chern number $C$\cite{Wan2011}, acting as the the monopole source and sink of the Berry curvature,
which leads to the striking predicitons of the Fermi arc electronic states on the surface\cite{Weng2015, Huang2015, Lv2015TaAs, Xu2015TaAs, Wang2015MoTe2, Sun2015MoTe2, Soluyanov2015WTe2, jiang2017signature}.

Macroscopically,  WPs are known to correlate to many exotic phenomena, such as negative magnetoresistance (also known as chiral anomaly)\cite{Xu2015TaAs, Xu2015NbAs}, intrinsic spin Hall effect\cite{Sun2016}, and anomalous Hall effect\cite{Burkov:2011de, Xu2011, Zhang_2017}.
Notably, if WPs with opposite chiral charges have different energies the quantized circular photogalvanic effect (CPGE) \cite{de2017quantized} and the chiral magnetic effect (CME) are also supposed to happen\cite{PhysRevD.78.074033, PhysRevB.89.035142, PhysRevB.92.161110, PhysRevB.91.115203, PhysRevB.88.125105, PhysRevLett.111.027201}. The CME has attracted much theoretical interest because of the existence of a nonzero chiral current induced by a static magnetic field at zero temperature\cite{PhysRevB.89.035142}. The CME and CPGE were first proposed by theoretical study and was recently observed in RhSi~\cite{RhSiCPGE}. Early theoretically proposed material candidates for WSMs have mirror symmetry, which confines the WPs with opposit chiral charges to the same energy.
Some recently proposed materials, which do not have mirror symmetry, have the WPs with higher chiral charges ($\geq 1$), such as the double Weyl fermions in SrSi$_2$\cite{Huang02022016}, TlTeO$_6$ and Ag$_2$Se related families\cite{chang2016theoretical}, and the spin-3/2 Rarita-Schwinger-Weyl fermions in CoSi\cite{Bradlynaaf5037, PhysRevLett.119.206402, pshenay2017band, chang2017large}. Therefore, a Weyl semimetal  with opposite unit chiral charges being at different energies is highly demanded, which would provide an ideal experimental platform facilitating  further examination of those exotic effects in the near future.

In this paper, based on the first-principle calculations and simple symmetry argument, we predict Ag$_2$Se at low temperature with space group (${\cal SG}$) 17 as a Weyl semimetal lacking of the protection of mirror symmetry. As a result, the WPs with opposite chirality charge reside at different energies. We found 24 WPs in this material with 20 of WPs locate below Fermi level($E_F$).
By tuning the energy relative to $E_F$, this system can serve as an ideal platform to study the significant CPGE and CME in experiments.
The persistent spin current discovered at $k_{z}=\pi$ surface further makes this system a good material candidate for spintronics. 

\section{Crystal structure and symmetry.}
At atmospheric pressure, the Ag$_2$Se crystallizes in two phases.
At high temperature Ag$_{2}$Se crystallizes in the bcc cubic phase (Im$\bar3m$)\cite{Ag2Se_cubic, Ag2Se_review}.
At 406~K a structure transition to the orthorhombic phase occurs,
where Ag$_2$Se can have two different structures with ${\cal SG}$~17 ($P222_1$)\cite{Ag2Se_SG17_1, Ag2Se_review}. and ${\cal SG}$~19 ($P2_12_12_1$)\cite{Ag2Se_SG19_1, Ag2Se_SG19_2, Ag2Se_review}.
The topological properties of ${\cal SG}$~19 phase Ag$_2$Se has been examined\cite{Ag2Se_SG19_topo}.
Here, we studied the ${\cal SG}$~17 phase whose topological nature, to the best of our knowledge, has not been explored.

Beside identity, ${\cal SG}$~17 contains three symmetries: a 2-fold rotation symmetry about $x$ axis, {\it i.e.} $C_{2x}=\{2_{100}|0,0,0\}$, and two screw rotation symmetries about $y$ and $z$ axis, {\it i.e.} $S_{2y}=\{2_{010}|0,0,\frac{1}{2}\}$ and $S_{2z}=\{2_{001}|0,0,\frac{1}{2}\}$.
Their operations are expressed as:
$C_{2x}:~(x, y, z)\rightarrow (x, -y, -z)\otimes(-i\sigma_{x}) $,
$S_{2y}:~(x, y, z)\rightarrow (-x, y, -z+c/2)\otimes(-i\sigma_{y})$,
$S_{2z}:~(x, y, z)\rightarrow (-x, -y, z+c/2)\otimes(-i\sigma_{z})$,
where the spin manipulation is indicated by the corresponding Pauli matrices.
Due to the absence of inversion symmetry and the Kramer degeneracy, at generic momentum every Bloch band  is non-degenerate.
However, higher band degeneracy can still be expected in this ${\cal SG}$: (1) each band at the eight time-reversal invariant momenta (TRIM) are doubly degenerate protected by the time-reversal symmetry ${\cal T}$, {\it i.e.} $|\Psi\rangle$ and ${\cal T}|\Psi\rangle$ are orthogonal and energetically degenerate at TRIM.
(2) Screw symmetry $S_{2z}$ protects crossings between two bands carrying different $S_{2z}$ eigenvalues.
It is easy to understand that, after applying $S_{2z}$ twice, every atom moves along $z$ axis with one lattice spacing, resulting in an additional phase factor in the Bloch wavefunction, {\it i.e.} $S_{2z}^{2}|\Psi(k)\rangle = -e^{ik_{z}}|\Psi(k)\rangle$ with the minus sign from the spin operation.
As $S_{2z}$ commutes with the Hamiltonian, each Bloch state can be labelled by the eigenvalue of $S_{2z}$ as well.
Although $k_{z}=\pm\pi$ are equivalent Brillouin Zone (BZ) boundaries, the eigenvalue of $S_{2z}$ takes different values, {\it i.e.} $\pm ie^{ik_{z}/2}=\pm1$ at $k_{z}=-\pi$ and $\pm ie^{ik_{z}/2}=\mp1$ at $k_{z}=\pi$.
Thus, to satisfy the periodic boundary condition of the BZ the bands with opposite $S_{2z}$ eigenvalues shall always appear in pairs~\cite{PhysRevLett.115.126803} and cross at somewhere along the high symmetry lines protected by $S_{2z}$. There are four k-paths of this type protected with band crossings, {\it i.e.} $(0/\pi, 0/\pi, k_{z})$.

(3) At $k_{z}=\pi$ plane every band are doubly degenerate due to the protection of the joint protection of ${\cal T}$ and $S_{2z}$, {\it i.e.} there exists a nodal plane in ${\cal SG}~17$ as shown below. After defining $\Theta_{z} = {\cal T}S_{2z}$, one can easily prove that the momentum at $k_{z}=0/\pi$ planes are invariant under $\Theta_{z}$,
$(k_{x}, k_{y}, 0/\pi) \xrightarrow{S_{2z}} (-k_{x}, -k_{y}, 0/\pi) \xrightarrow{\cal T} (k_{x}, k_{y}, 0/\pi).$
While, in real-space after applying $\Theta_{z}$ twice the $x$ and $y$ coordinates are kept invariant, but $z$ coordinate will be shifted by one lattice spacing $c$,
$(x, y, z)\xrightarrow{S_{2z}}(-x, -y, z+c/2)\xrightarrow{\cal T}(-x, -y, z+c/2).$
As a result, for each Bloch band $\Theta_{z}^{2}|\Psi(k)\rangle = e^{ik_{z}}|\Psi(k)\rangle$.
At $k_{z}=\pi$ plane, $\Theta_{z}^{2}=-1$ acting as a time-reversal operator which allows a double band degeneracy everywhere in this plane.
The other screw rotation, {\it i.e.} $S_{2y}$, however does not bring in any additional band degeneracy, as applying $\Theta_{y} = {\cal T}S_{2y}$ twice will not result in additional lattice translation so that $\Theta_{y}^{2} = 1$.
\begin{figure}[t]
\centering
\includegraphics[width=0.95\linewidth]{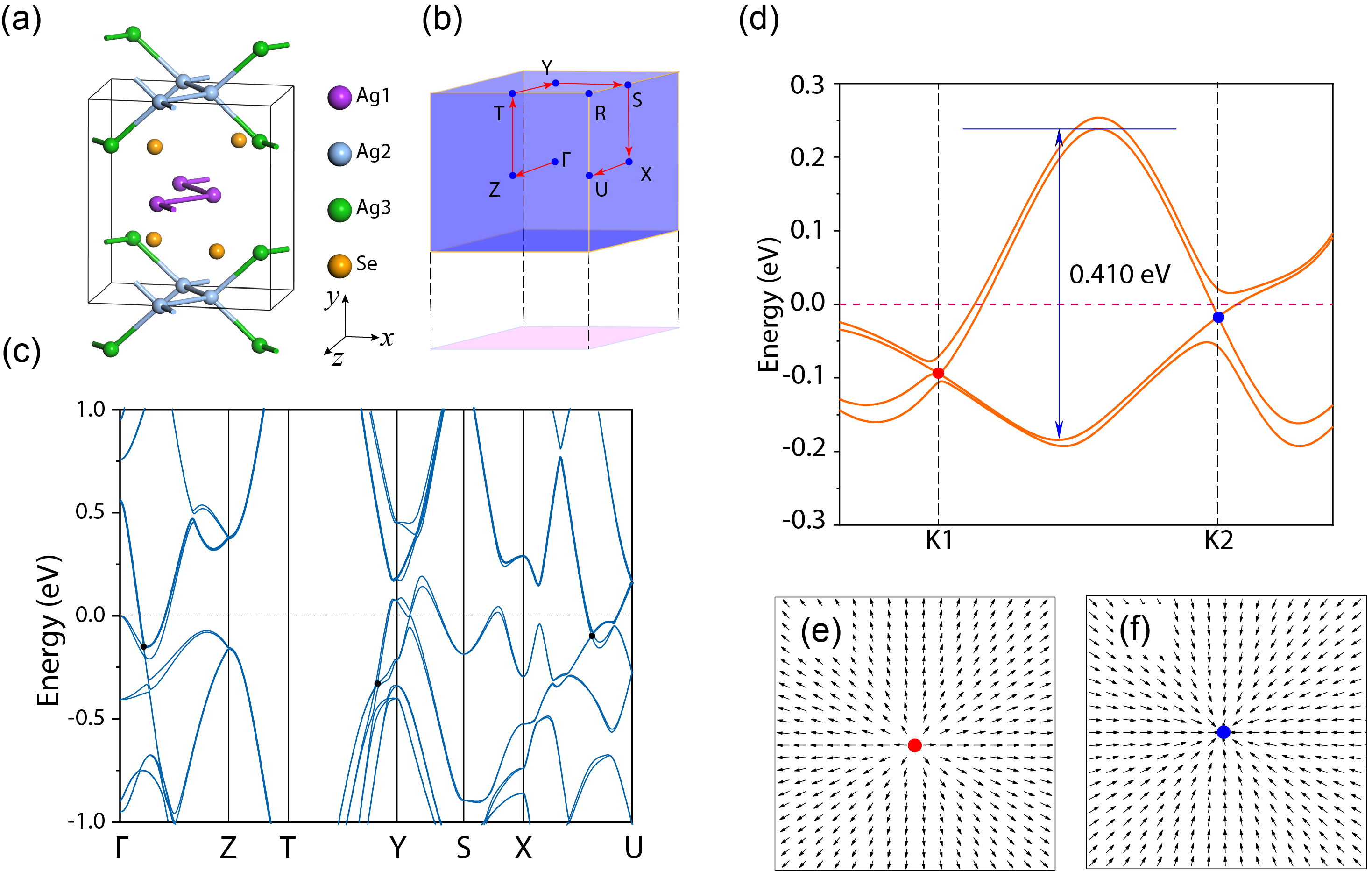}
\caption{
The lattice structure of the Ag$_2$Se (a) and its BZ (b) with the (010) projection. The bulk band structure with SOC is shown in (c) along high-symmetry path indicated in (b). (d) illustrates a pair of WPs K1 (0.500 0.000 -0.315), K2 (0.195 0.000 -0.153) in the unit $(\frac{2\pi}{a}, \frac{2\pi}{b}, \frac{2\pi}{c})$ and the two associated bands inverted in between with the inversion energy 0.41 eV. (e) and (f) demonstrate the berry curvature associated with these two WPS. In (c), the black dots indicate the band cross, and the dashed line indicates the Fermi level. 
}
\label{fig:structure}
\end{figure}

(4) What is more striking about ${\cal SG}$~17 is the protection of nonequilibrium spin polarization along the boundary of $k_{z}=\pi$ surface, which strongly connects to the nodal nature of this plane, resulting in a persistent spin texture allowed only in selective directions in momentum space.
As $\Theta_{z}|\Psi\rangle$ is energetically degenerate with $|\Psi\rangle$, it is easy to verify $s_{z}=\langle\Psi|\Theta_{z}^{\dagger}~\hat{\sigma}_{z}~\Theta_{z}|\Psi\rangle = -\langle \Psi| \hat{\sigma}_{z}|\Psi\rangle$ due to the fact that $\hat{\sigma}_{z}$ anticommutes with ${\cal T}$ and commutes with $S_{2z}$.
Adding up the contributions from the two degenerate states yields a vanishing spin polarization along $z$ direction.
However, the other two spin components survive as $\{S_{2z}, \hat{\sigma}_{x/y}\}=0$.
We note that this strongly contrasts to the spin textures in centrosymmetric systems, where the Kramer's degeneracy enforces the doubly degenerate bands to have perfectly opposite spin textures.
Therefore a cancellation of spin polarization happens everywhere in the BZ.
Here the $\Theta_{z}$ protected nodal plane promotes the cancellation only for the $\langle S_{z}\rangle$ component, leaving the spin momentum fully planar.

\section{bulk electronic structure}
For systems in this ${\cal SG}$ the more interesting thing is the presence of WPs at generic momentum due to the broken of inversion symmetry, which cannot be obtained from the above symmetry analysis.
To this end, we calculated the electronic band structure and search the possible WPs in the first BZ by using the density functional theory (DFT) as implemented in the Vienna {\it ab-initio} Simulation Package (VASP)\cite{vasp} with projector augmented wave method \cite{paw1, paw2}.
The electron exchange-correlation potential is described by the generalized gradient approximation of Perdew-Burke-Ernzerhof (GGA-PBE) scheme\cite{GGA}. The kinetic energy cutoff of the plane wave basis is set to 300~eV as default. A $\Gamma$-centered $6\times6\times10$ $k$-mesh is used to sample the BZ.
To calculate the surface states, the surface Green's function method was employed with the tight-binding Hamiltonian constructed by the maximally localized Wannier orbitals\cite{mostofi2008wannier90, MOSTOFI20142309}, where the Se $s$, $p$; and Ag $s$, $p$, and $d$ orbitals were considered.
The experiment lattice parameters ($a=7.050$~\AA, $b=7.850$~\AA, and $c=4.330$~\AA) were employed in the calculations with the crystal structure shown in Fig.~\ref{fig:structure}(a).
The BZ and the corresponding (010) surface BZ are displayed in Fig.~\ref{fig:structure}(b).
The electronic structure calculated with spin-orbit coupling (SOC) along the high symmetry path is shown in Fig.~\ref{fig:structure}(c).

\begin{figure}[htbp]
\centering
\includegraphics[width=0.95\linewidth]{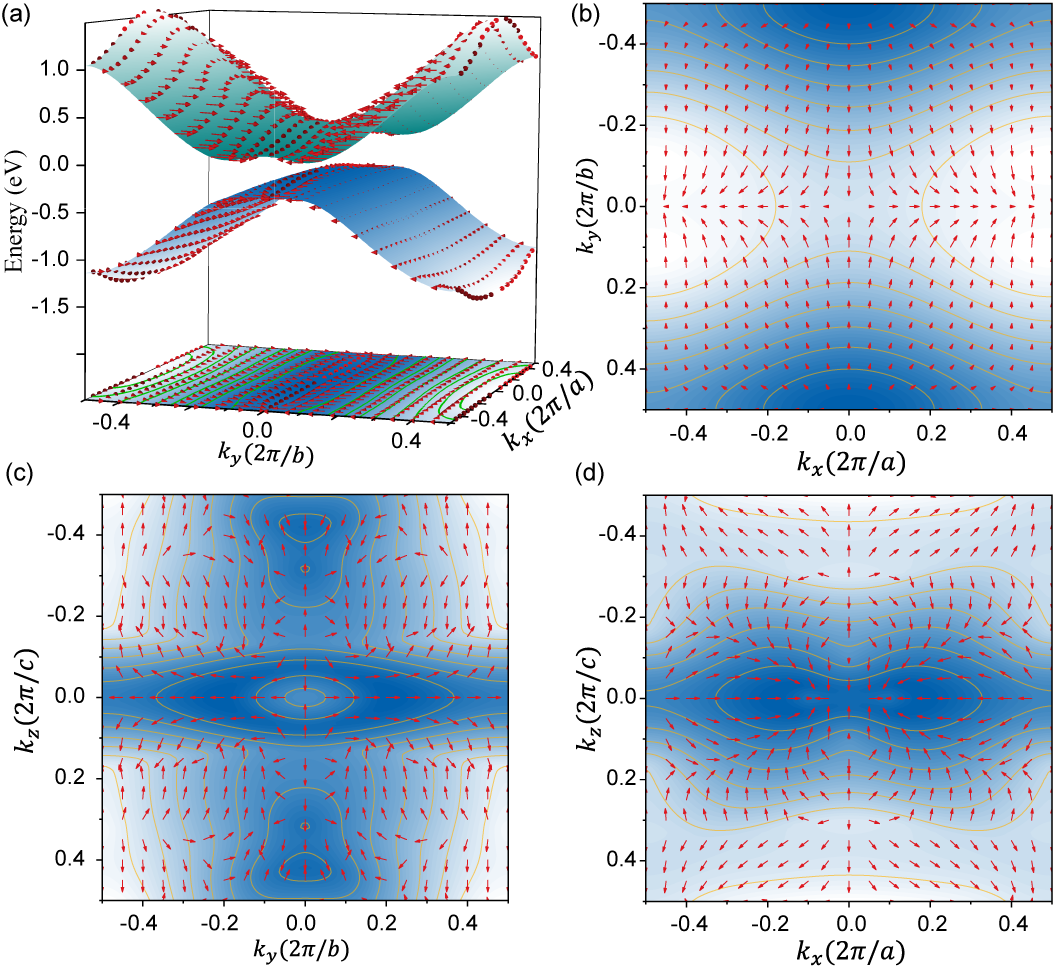}
\caption{
Spin texture of the top valence band at (a, b) $k_{z}=\pi$; (c) $k_{x}=\pi$ and (d) $k_{y}=\pi$ planes. In addition, in (a) the spin texture of the bottom conduction band is also shown and further overlaid on the band surface.
}
\label{fig:spin-texture}
\end{figure}

\begin{figure*}[htbp]
\centering
\includegraphics[width=0.9\textwidth]{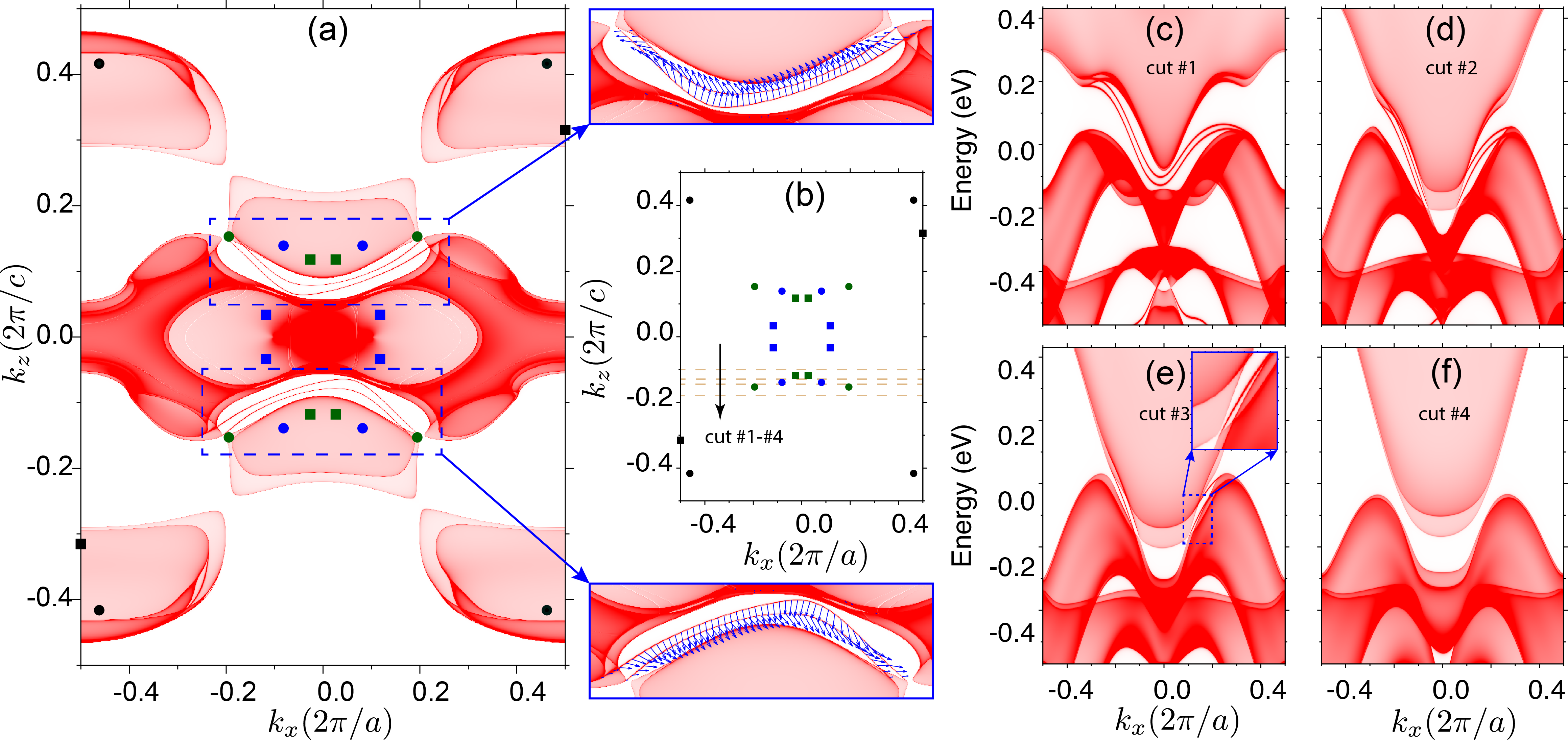}
\caption{
(a) The calculated surface states at the Fermi level and the (010) surface with the Fermi arcs and their spin texture highlighted in the dashed line box. The location of the 24 WPs are shown in (b) with four momentum cuts, each of which separats the WPs into two parts with different net Chern numbers. Along the four momentum cuts, the corresponding electronic states are shown in (c-f) with different number of surface states highly consistent with the prediction of the enclosed Chern number by the momentum cut. In (a) and (b), the filled squares and circles indicate the WPs with opposite chiral charges and different color indicate the different groups, and the black solid arrows indicate two WPs projected together with the net chiral charges of $|C| = 2$, while other WPs without arrows indicates the net chiral charges of $|C| = 1$.
}
\label{fig:fs}
\end{figure*}

Bloch states along $\Gamma$-Z, T-Y and X-U are the eigenstates of $S_{2z}$ screw symmetry, therefore there will be $S_{2z}$ enforced band crossings at these three lines as explained before.
Along each line we marked one such crossing point by black dot and we note that each band will at least have one crossing point with another band carrying different $S_{2z}$ eigenvalues.
However, their locations in momentum-energy space are unpredictable by symmetry and is material-dependent.
In addition to the $S_{2z}$ symmetry-enforced band crossing at high-symmetry lines confirmed by our DFT calculations, the $\Theta_{z}={\cal T}S_{2z}$ protected nodal plane ($k_{z}=\pi$) was also verified numerically, where every band is confirmed to be doubly degenerate.

In Fig.~\ref{fig:spin-texture}(a) we show the bottom conduction band and the top valence band at $k_{z}=\pi$ plane, where each band is doubly degenerate and spin momentum contains only in-plane components, {\it i.e.} $\langle S_{x}\rangle$ and $\langle S_{y}\rangle$.
For a better visualization, the spin texture for the valence band is projected to the $k_{x}-k_{z}$ plane.
We first note that, along $k_{y}=0,\pi$ lines, only the $\langle S_{y}\rangle$ component is nonzero.
This is due to the additional constraint from $S_{2y}$ which keeps the momentum invariant along these lines and it commute with the Hamiltonian.
Each eigenstate of the Hamiltonian, thus, will carry a definite eigenvalue of $S_{2y}$ as well.
From $S_{2y}^{2} = -1$, the eigenvalue of $S_{2y}$ can be either $i$ or $-i$, which classifies the Bloch states into two groups.
For a given state $|\Psi^{(+)}\rangle$, for example with $S_{2y}$ eigenvalue $i$, the anti-commutation relation $\{\sigma_{y}, \sigma_{x}\} =0$ results in $\langle S_{x}\rangle\sim\langle \Psi^{(+)}| \sigma_{x}|\Psi^{(+)}\rangle=\langle \Psi^{(+)}| S_{2y}^{\dagger} \sigma_{x} S_{2y}|\Psi^{(+)}\rangle = -\langle \Psi^{(+)}| \sigma_{x}|\Psi^{(+)}\rangle=0$.
As a consequence, along $k_{y}=0/\pi$ lines at $k_{z}=\pi$ plane the only non-zero component of the spin is $\langle S_{y}\rangle$, which forces the spin transports only along $k_{y}$ direction.
As displayed in Fig.~\ref{fig:spin-texture}(b), the spin texture along these lines are perfectly aligning to $k_{y}$ direction only.
The transport of spin current would then be protected by the symmetry against impurities and defects scattering.
More interestingly, we find that, in Ag$_{2}$Se, the persistent of spin texture extends to a large regime around these lines, making the majority of spin texture at $k_{z}=\pi$ plane orientated to $k_{y}$.
The persistent spin texture has been predicted to support long spin lifetime and create non-equilibrium spin polarization~\cite{EDELSTEIN1990233} which is crucial for spintronic applications~\cite{Tao2018}.
In the other two planes, {\it i.e.} $k_{x}=\pi$ in Fig.~\ref{fig:spin-texture}(c) and $k_{y}=\pi$ in Fig.~\ref{fig:spin-texture}(d), the spin textures are three dimensional and are much more random than the $k_{z}=\pi$ plane.

\section{Weyl points, surface states and qpi}
Furthermore, we found 24 WPs around the Fermi level, see Table~\ref{tb:wps} for their positions in momentum-energy space.
These WPs are symmetry related.
Given a WP at an arbitrary $k$ point ($k_x, k_y, k_z$), $e.g.$, $C_{2x}$ rotation symmetry and $\mathcal {T}$ guarantee that there will be three more WPs locating at ($k_x, -k_y, -k_z$), ($-k_x, -k_y, -k_z$), and ($-k_x, k_y, k_z$).
As neither $C_{2}$ and ${\cal T}$ changes Chern number, all these symmetry-connected WPs have the same chirality. Therefore, the 24 WPs can be devided into six groups. Within each group, the WPs are related by ${\cal T}$ and $C_{2}$.
 However, WPs from different groups have different energies due to the broken of mirror symmetry.
Thus, a net chirality charge can be achieved by tuning the Fermi level in this system, which has a strong implication to the surface states and the realization of CME and CPGE at low temperature.
\begin{table}[t]
  \caption{The six groups of ${\cal SG}$~17 Ag$_2$Se WPs. The WP position (in reduced coordinates $x, y, z$), Chern number, and the energy relative to the Fermi energy ($E_F$). For each group, there are four symmetry-related WPs. }
  \label{tb:wps}
  \begin{tabular}{ccccc}
  \hline
  \hline
      WPs    & Coordinates                                     &$E-E_F$& Chern \\
             &($\frac{2\pi}{a},\frac{2\pi}{a},\frac{2\pi}{c}$) & (eV)  & number\\
             \hline
  $W_1^+$    &(0.082, 0.000, 0.139) &-0.191 & $+1$\\
  $W_2^+$    &(0.195, 0.000, 0.153) &-0.017 & $+1$\\
  $W_3^+$    &(0.462, 0.000, 0.416) &-0.043 & $+1$\\
  $W_4^-$    &(0.026, 0.000, 0.118) &-0.199 & $-1$\\
  $W_5^-$    &(0.118, 0.500, 0.034) & 0.200 & $-1$\\
  $W_6^-$    &(0.500, 0.000, 0.315) &-0.094 & $-1$\\
  \hline
  \hline
  \end{tabular}
\end{table}
To see clearly the energy difference between a pair of WPs, we show in Fig.~\ref{fig:structure}(d) the bands connecting a chosen pair of WPs.
The location of two WPs ($K_{1}$ and $K_{2}$) are denoted in this figure by the red and blue dots.
The energy separation of the two WPs is as large as 70 meV.
We also notice that the two bands forming this pair of WPs exchange orders between $K_{1}$ and $K_{2}$ resulting in a band inversion, whose strength is characterized by an energy gap of 0.410~eV.
 In Fig.~\ref{fig:structure}~(e) and (f) the Berry curvature around this pair of WPs demonstrate clear opposite chirality.

The characteristic feature of Weyl systems is the charge transport between the Weyl points through the Fermi arc.
In Fig.~\ref{fig:fs} the states at the Fermi level are displayed for (010) surface.
We find four arcs as highlighted inside the dashed line box where their spin textures are also displayed.
As the WPs reside at different energies, it is impossible to get a pair of WPs on the same energy cut.
Instead, one can inspect the relation of the Fermi arcs with the Chern number by introducing momentum cut.
In Fig.~\ref{fig:fs}(b) we show the locations of the 24 WPs in the surface BZ.
We use square and circle to denote the positive and negative chirality.
The different colors represent the different groups.
We note that some WPs are not visible in this plot as they stay vertically below/above some of these WPs in energy.
As shown in Fig.~\ref{fig:fs}(b), we introduce four momentum cuts along $k_x$ axis.
Each momentum cut separates the BZ into two parts, each of which encloses different number of WPs yielding a net Chern number.
Consequently, there will be surface states passing through this momentum cut if the enclosed Chern number is nonzero.
The number of the topological surface states must be same as the enclosed Chern number.
The calculated results are displayed in Fig.~\ref{fig:fs}(c-f). For cut \#1, the enclosed Chern number is $+2$, therefore, will be 2 surface states. For cut \#2, the there are four topological surface states, it is consistent with the enclosed Chern number $+4$, while for cut \#3, the Chern number decreases to $+2$ because of two more $-2$ Chern number WPs are enclosed, there are two topological surface states. For cut \#4, the enclosed Chern number is zero. Consequently, there is no topological surface state at all passing through this momentum cut.

\begin{figure}[htbp]
\centering
\includegraphics[width=0.95\linewidth]{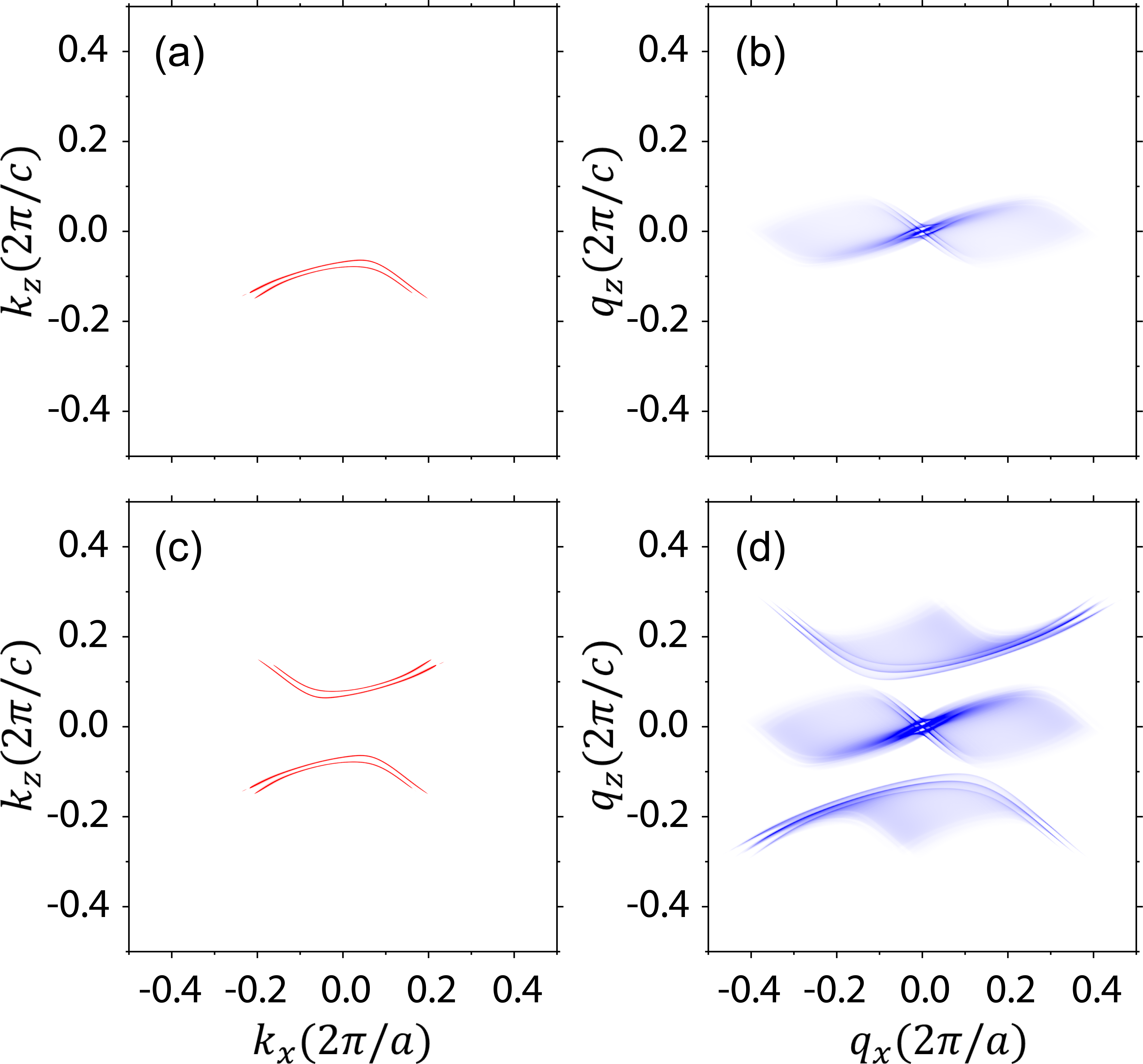}
\caption{Quasiparticle interference pattern induced by the Fermi arc states shown in Fig.~\ref{fig:fs}(a). One branch of the fermi arc (a) leads to the scattering among the states within it, whose QPI (b) is distributed only around $\Gamma$ point.  The scattering between the two branches of the Fermi arcs shown in (c) leads to the additional QPI mode locating at higher $q_{z}$ value (d). }
\label{fig:qpi}
\end{figure}

Through both energy- and momentum-cut we have verified that the discovered 24 WPs are indeed topological and Ag$_{2}$Se in ${\cal SG}$~17 is a Weyl semimetal holding great potential for CME and CPGE.
To help experimental characterization and verification of this topologically nontrivial system, we further calculate the quasiparticle interference (QPI) pattern which can be directly probed by scanning tunneling spectroscopy (STM).
QPI measures the scattering pattern between the initial surface states at $k_i$, and the final states at $k_f$ at a fixed energy,
which is theoretically represented by the joint density of states (JDOS)~\cite{Kourtis2016}
$\text{JDOS}(q,E) = \int\ \frac{d^{3}k}{(2\pi)^3}\ \sum_\sigma A_{\sigma\sigma}(k,E)\sum_{\sigma'} A_{\sigma'\sigma'}(k+q,E)$,
where $A_{\sigma\sigma}(k,E)\sum_{\sigma'}$ is the Fourier-transformed density of states with spin $\sigma (\sigma')$ at $k$ point and energy $E$, which can be obtain from the imaginary part of the surface Green's function.
To trace the scattering pattern resulted solely by the topological nature of this system, we only consider the QPI between the surface Fermi arc shown in Fig.~\ref{fig:fs}~(a).
For a better understanding, we dissemble the Fermi arc shown in Fig.~\ref{fig:fs} and calculate the JDOS associated with them.
With only one branch of the Fermi arcs, scattering is allowed only between the states within this arc.
As shown in Fig.~\ref{fig:qpi}(b) the Fermi arc in Fig.~\ref{fig:qpi} results in a bow tie shape QPI distributing around the BZ center.
With the inclusion of the other branch shown in Fig.~\ref{fig:qpi}(c), the scattering between the two Fermi arcs is now allowed, which additionally leads to the QPI patten locating above/below the bow tie QPI.
The entire QPI spans over a large territory of the surface BZ, which provides a great opportunity for STM to detect it.
We hope this would stimulate interest for relevant experimental techniques.


\section{Conclusion}
We have studied the ${\cal SG}$~17 whose symmetry operation is simple but astonishingly rich phases can be derived which have strong implication to many exciting phenomena.
We found a symmetry protected nodal plane in this noncentrosymmetric system protected by the joint operation of time-reversal ${\cal T}$ and screw rotation $S_{2z}$ symmetries.
Along the edge of the nodal plane, a direction selective spin polarization is achieved. Unlike the quantum spin Hall system, the polarization of the spin current does not require the presence of perfectly long edge which is technically hard to achieve.
The direction selective spin polarization is a bulk quantity determined completely by the symmetry.
The presence of persistent spin texture in momentum space may utilize potential applications in spintronics.
By using the first-principle calculations, we also found 24 WPs locating at generic momentum with notable Ferm arcs and large QPI.
The absence of mirror symmetry makes this system also a perfect material platform for quantized circular photogalvanic and the chiral magnetic effect.


\begin{acknowledgments}
W.S. would like to thank Max Planck Institute for Chemical Physics of Solids at Dresden for the hospitality where part of this work was done. W.S. wants to acknowledge Yan Sun and Claudia Felser for the fruitful collaborations in other projects.
G.L. wants to thank Shilei Zhang for helpful discussions on the momentum-space Skyrmion texture.
G.L. acknowledges the starting grant of ShanghaiTech University and the Program for Professor of Special Appointment (Shanghai Eastern Scholar).
Calculations were partly carried out at the HPC Platform of ShanghaiTech University Library and Information Services, as well as School of Physical Science and Technology.
\end{acknowledgments}

\bibliography{TopMater}

\begin{thebibliography}{47}%
\makeatletter
\providecommand \@ifxundefined [1]{%
 \@ifx{#1\undefined}
}%
\providecommand \@ifnum [1]{%
 \ifnum #1\expandafter \@firstoftwo
 \else \expandafter \@secondoftwo
 \fi
}%
\providecommand \@ifx [1]{%
 \ifx #1\expandafter \@firstoftwo
 \else \expandafter \@secondoftwo
 \fi
}%
\providecommand \natexlab [1]{#1}%
\providecommand \enquote  [1]{``#1''}%
\providecommand \bibnamefont  [1]{#1}%
\providecommand \bibfnamefont [1]{#1}%
\providecommand \citenamefont [1]{#1}%
\providecommand \href@noop [0]{\@secondoftwo}%
\providecommand \href [0]{\begingroup \@sanitize@url \@href}%
\providecommand \@href[1]{\@@startlink{#1}\@@href}%
\providecommand \@@href[1]{\endgroup#1\@@endlink}%
\providecommand \@sanitize@url [0]{\catcode `\\12\catcode `\$12\catcode
  `\&12\catcode `\#12\catcode `\^12\catcode `\_12\catcode `\%12\relax}%
\providecommand \@@startlink[1]{}%
\providecommand \@@endlink[0]{}%
\providecommand \url  [0]{\begingroup\@sanitize@url \@url }%
\providecommand \@url [1]{\endgroup\@href {#1}{\urlprefix }}%
\providecommand \urlprefix  [0]{URL }%
\providecommand \Eprint [0]{\href }%
\providecommand \doibase [0]{http://dx.doi.org/}%
\providecommand \selectlanguage [0]{\@gobble}%
\providecommand \bibinfo  [0]{\@secondoftwo}%
\providecommand \bibfield  [0]{\@secondoftwo}%
\providecommand \translation [1]{[#1]}%
\providecommand \BibitemOpen [0]{}%
\providecommand \bibitemStop [0]{}%
\providecommand \bibitemNoStop [0]{.\EOS\space}%
\providecommand \EOS [0]{\spacefactor3000\relax}%
\providecommand \BibitemShut  [1]{\csname bibitem#1\endcsname}%
\let\auto@bib@innerbib\@empty
\bibitem [{\citenamefont {Wan}\ \emph {et~al.}(2011)\citenamefont {Wan},
  \citenamefont {Turner}, \citenamefont {Vishwanath},\ and\ \citenamefont
  {Savrasov}}]{Wan2011}%
  \BibitemOpen
  \bibfield  {author} {\bibinfo {author} {\bibfnamefont {Xian~Gang}\
  \bibnamefont {Wan}}, \bibinfo {author} {\bibfnamefont {Ari~M}\ \bibnamefont
  {Turner}}, \bibinfo {author} {\bibfnamefont {Ashvin}\ \bibnamefont
  {Vishwanath}}, \ and\ \bibinfo {author} {\bibfnamefont {Sergey~Y}\
  \bibnamefont {Savrasov}},\ }\bibfield  {title} {\enquote {\bibinfo {title}
  {{Topological semimetal and Fermi-arc surface states in the electronic
  structure of pyrochlore iridates}},}\ }\href@noop {} {\bibfield  {journal}
  {\bibinfo  {journal} {Phys. Rev. B}\ }\textbf {\bibinfo {volume} {83}},\
  \bibinfo {pages} {205101} (\bibinfo {year} {2011})}\BibitemShut {NoStop}%
\bibitem [{\citenamefont {Weng}\ \emph {et~al.}(2015)\citenamefont {Weng},
  \citenamefont {Fang}, \citenamefont {Fang}, \citenamefont {Bernevig},\ and\
  \citenamefont {Dai}}]{Weng2015}%
  \BibitemOpen
  \bibfield  {author} {\bibinfo {author} {\bibfnamefont {Hongming}\
  \bibnamefont {Weng}}, \bibinfo {author} {\bibfnamefont {Chen}\ \bibnamefont
  {Fang}}, \bibinfo {author} {\bibfnamefont {Zhong}\ \bibnamefont {Fang}},
  \bibinfo {author} {\bibfnamefont {B.~Andrei}\ \bibnamefont {Bernevig}}, \
  and\ \bibinfo {author} {\bibfnamefont {Xi}~\bibnamefont {Dai}},\ }\bibfield
  {title} {\enquote {\bibinfo {title} {{Weyl Semimetal Phase in
  Noncentrosymmetric Transition-Metal Monophosphides}},}\ }\href@noop {}
  {\bibfield  {journal} {\bibinfo  {journal} {Phys. Rev. X}\ }\textbf {\bibinfo
  {volume} {5}},\ \bibinfo {pages} {011029} (\bibinfo {year}
  {2015})}\BibitemShut {NoStop}%
\bibitem [{\citenamefont {Huang}\ \emph {et~al.}(2015)\citenamefont {Huang},
  \citenamefont {Xu}, \citenamefont {Belopolski}, \citenamefont {Lee},
  \citenamefont {Chang}, \citenamefont {Wang}, \citenamefont {Alidoust},
  \citenamefont {Bian}, \citenamefont {Neupane}, \citenamefont {Zhang},
  \citenamefont {Jia}, \citenamefont {Bansil}, \citenamefont {Lin},\ and\
  \citenamefont {Hasan}}]{Huang2015}%
  \BibitemOpen
  \bibfield  {author} {\bibinfo {author} {\bibfnamefont {Shin-Ming}\
  \bibnamefont {Huang}}, \bibinfo {author} {\bibfnamefont {Su-Yang}\
  \bibnamefont {Xu}}, \bibinfo {author} {\bibfnamefont {Ilya}\ \bibnamefont
  {Belopolski}}, \bibinfo {author} {\bibfnamefont {Chi-Cheng}\ \bibnamefont
  {Lee}}, \bibinfo {author} {\bibfnamefont {Guoqing}\ \bibnamefont {Chang}},
  \bibinfo {author} {\bibfnamefont {BaoKai}\ \bibnamefont {Wang}}, \bibinfo
  {author} {\bibfnamefont {Nasser}\ \bibnamefont {Alidoust}}, \bibinfo {author}
  {\bibfnamefont {Guang}\ \bibnamefont {Bian}}, \bibinfo {author}
  {\bibfnamefont {Madhab}\ \bibnamefont {Neupane}}, \bibinfo {author}
  {\bibfnamefont {Chenglong}\ \bibnamefont {Zhang}}, \bibinfo {author}
  {\bibfnamefont {Shuang}\ \bibnamefont {Jia}}, \bibinfo {author}
  {\bibfnamefont {Arun}\ \bibnamefont {Bansil}}, \bibinfo {author}
  {\bibfnamefont {Hsin}\ \bibnamefont {Lin}}, \ and\ \bibinfo {author}
  {\bibfnamefont {M~Z}\ \bibnamefont {Hasan}},\ }\bibfield  {title} {\enquote
  {\bibinfo {title} {{A Weyl Fermion semimetal with surface Fermi arcs in the
  transition metal monopnictide TaAs class}},}\ }\href@noop {} {\bibfield
  {journal} {\bibinfo  {journal} {Nat. Commun.}\ }\textbf {\bibinfo {volume}
  {6}},\ \bibinfo {pages} {8373} (\bibinfo {year} {2015})}\BibitemShut
  {NoStop}%
\bibitem [{\citenamefont {Lv}\ \emph {et~al.}(2015)\citenamefont {Lv},
  \citenamefont {Weng}, \citenamefont {Fu}, \citenamefont {Wang}, \citenamefont
  {Miao}, \citenamefont {Ma}, \citenamefont {Richard}, \citenamefont {Huang},
  \citenamefont {Zhao}, \citenamefont {Chen}, \citenamefont {Fang},
  \citenamefont {Dai}, \citenamefont {Qian},\ and\ \citenamefont
  {Ding}}]{Lv2015TaAs}%
  \BibitemOpen
  \bibfield  {author} {\bibinfo {author} {\bibfnamefont {B~Q}\ \bibnamefont
  {Lv}}, \bibinfo {author} {\bibfnamefont {H~M}\ \bibnamefont {Weng}}, \bibinfo
  {author} {\bibfnamefont {B~B}\ \bibnamefont {Fu}}, \bibinfo {author}
  {\bibfnamefont {X~P}\ \bibnamefont {Wang}}, \bibinfo {author} {\bibfnamefont
  {H}~\bibnamefont {Miao}}, \bibinfo {author} {\bibfnamefont {J}~\bibnamefont
  {Ma}}, \bibinfo {author} {\bibfnamefont {P}~\bibnamefont {Richard}}, \bibinfo
  {author} {\bibfnamefont {X~C}\ \bibnamefont {Huang}}, \bibinfo {author}
  {\bibfnamefont {L~X}\ \bibnamefont {Zhao}}, \bibinfo {author} {\bibfnamefont
  {G~F}\ \bibnamefont {Chen}}, \bibinfo {author} {\bibfnamefont
  {Z.}~\bibnamefont {Fang}}, \bibinfo {author} {\bibfnamefont {X.}~\bibnamefont
  {Dai}}, \bibinfo {author} {\bibfnamefont {T}~\bibnamefont {Qian}}, \ and\
  \bibinfo {author} {\bibfnamefont {H}~\bibnamefont {Ding}},\ }\bibfield
  {title} {\enquote {\bibinfo {title} {{Experimental Discovery of Weyl
  Semimetal TaAs}},}\ }\href@noop {} {\bibfield  {journal} {\bibinfo  {journal}
  {Phys. Rev. X}\ }\textbf {\bibinfo {volume} {5}},\ \bibinfo {pages} {031013}
  (\bibinfo {year} {2015})}\BibitemShut {NoStop}%
\bibitem [{\citenamefont {Xu}\ \emph {et~al.}(2015{\natexlab{a}})\citenamefont
  {Xu}, \citenamefont {Belopolski}, \citenamefont {Alidoust}, \citenamefont
  {Neupane}, \citenamefont {Bian}, \citenamefont {Zhang}, \citenamefont
  {Sankar}, \citenamefont {Chang}, \citenamefont {Zhujun}, \citenamefont {Lee},
  \citenamefont {Shin-Ming}, \citenamefont {Zheng}, \citenamefont {Ma},
  \citenamefont {Sanchez}, \citenamefont {Wang}, \citenamefont {Bansil},
  \citenamefont {Chou}, \citenamefont {Shibayev}, \citenamefont {Lin},
  \citenamefont {Jia},\ and\ \citenamefont {Hasan}}]{Xu2015TaAs}%
  \BibitemOpen
  \bibfield  {author} {\bibinfo {author} {\bibfnamefont {Su-Yang}\ \bibnamefont
  {Xu}}, \bibinfo {author} {\bibfnamefont {Ilya}\ \bibnamefont {Belopolski}},
  \bibinfo {author} {\bibfnamefont {Nasser}\ \bibnamefont {Alidoust}}, \bibinfo
  {author} {\bibfnamefont {Madhab}\ \bibnamefont {Neupane}}, \bibinfo {author}
  {\bibfnamefont {Guang}\ \bibnamefont {Bian}}, \bibinfo {author}
  {\bibfnamefont {Chenglong}\ \bibnamefont {Zhang}}, \bibinfo {author}
  {\bibfnamefont {Raman}\ \bibnamefont {Sankar}}, \bibinfo {author}
  {\bibfnamefont {Guoqing}\ \bibnamefont {Chang}}, \bibinfo {author}
  {\bibfnamefont {Yuan}\ \bibnamefont {Zhujun}}, \bibinfo {author}
  {\bibfnamefont {Chi-Cheng}\ \bibnamefont {Lee}}, \bibinfo {author}
  {\bibfnamefont {Huang}\ \bibnamefont {Shin-Ming}}, \bibinfo {author}
  {\bibfnamefont {Hao}\ \bibnamefont {Zheng}}, \bibinfo {author} {\bibfnamefont
  {Jie}\ \bibnamefont {Ma}}, \bibinfo {author} {\bibfnamefont {Daniel~S.}\
  \bibnamefont {Sanchez}}, \bibinfo {author} {\bibfnamefont {BaoKai}\
  \bibnamefont {Wang}}, \bibinfo {author} {\bibfnamefont {Arun}\ \bibnamefont
  {Bansil}}, \bibinfo {author} {\bibfnamefont {Fangcheng}\ \bibnamefont
  {Chou}}, \bibinfo {author} {\bibfnamefont {Pavel~P.}\ \bibnamefont
  {Shibayev}}, \bibinfo {author} {\bibfnamefont {Hsin}\ \bibnamefont {Lin}},
  \bibinfo {author} {\bibfnamefont {Shuang}\ \bibnamefont {Jia}}, \ and\
  \bibinfo {author} {\bibfnamefont {M.~Zahid}\ \bibnamefont {Hasan}},\
  }\bibfield  {title} {\enquote {\bibinfo {title} {{Discovery of a Weyl fermion
  semimetal and topological Fermi arcs}},}\ }\href@noop {} {\bibfield
  {journal} {\bibinfo  {journal} {Science}\ }\textbf {\bibinfo {volume}
  {349}},\ \bibinfo {pages} {613} (\bibinfo {year}
  {2015}{\natexlab{a}})}\BibitemShut {NoStop}%
\bibitem [{\citenamefont {Wang}\ \emph
  {et~al.}(2016{\natexlab{a}})\citenamefont {Wang}, \citenamefont {Gresch},
  \citenamefont {Soluyanov}, \citenamefont {Xie}, \citenamefont {Kushwaha},
  \citenamefont {Dai}, \citenamefont {Troyer}, \citenamefont {Cava},\ and\
  \citenamefont {Bernevig}}]{Wang2015MoTe2}%
  \BibitemOpen
  \bibfield  {author} {\bibinfo {author} {\bibfnamefont {Zhijun}\ \bibnamefont
  {Wang}}, \bibinfo {author} {\bibfnamefont {Dominik}\ \bibnamefont {Gresch}},
  \bibinfo {author} {\bibfnamefont {Alexey~A}\ \bibnamefont {Soluyanov}},
  \bibinfo {author} {\bibfnamefont {Weiwei}\ \bibnamefont {Xie}}, \bibinfo
  {author} {\bibfnamefont {S}~\bibnamefont {Kushwaha}}, \bibinfo {author}
  {\bibfnamefont {Xi}~\bibnamefont {Dai}}, \bibinfo {author} {\bibfnamefont
  {Matthias}\ \bibnamefont {Troyer}}, \bibinfo {author} {\bibfnamefont
  {Robert~J}\ \bibnamefont {Cava}}, \ and\ \bibinfo {author} {\bibfnamefont
  {B.~Andrei}\ \bibnamefont {Bernevig}},\ }\bibfield  {title} {\enquote
  {\bibinfo {title} {{MoTe2: A Type-II Weyl Topological Metal}},}\ }\href@noop
  {} {\bibfield  {journal} {\bibinfo  {journal} {Phys. Rev. Lett.}\ }\textbf
  {\bibinfo {volume} {117}},\ \bibinfo {pages} {056805} (\bibinfo {year}
  {2016}{\natexlab{a}})}\BibitemShut {NoStop}%
\bibitem [{\citenamefont {Sun}\ \emph {et~al.}(2015)\citenamefont {Sun},
  \citenamefont {Wu}, \citenamefont {Ali}, \citenamefont {Felser},\ and\
  \citenamefont {Yan}}]{Sun2015MoTe2}%
  \BibitemOpen
  \bibfield  {author} {\bibinfo {author} {\bibfnamefont {Y.}~\bibnamefont
  {Sun}}, \bibinfo {author} {\bibfnamefont {S~C}\ \bibnamefont {Wu}}, \bibinfo
  {author} {\bibfnamefont {M~N}\ \bibnamefont {Ali}}, \bibinfo {author}
  {\bibfnamefont {C.}~\bibnamefont {Felser}}, \ and\ \bibinfo {author}
  {\bibfnamefont {B}~\bibnamefont {Yan}},\ }\bibfield  {title} {\enquote
  {\bibinfo {title} {{Prediction of Weyl semimetal in orthorhombic MoTe 2}},}\
  }\href@noop {} {\bibfield  {journal} {\bibinfo  {journal} {Phys. Rev. B}\
  }\textbf {\bibinfo {volume} {92}},\ \bibinfo {pages} {161107(R)} (\bibinfo
  {year} {2015})}\BibitemShut {NoStop}%
\bibitem [{\citenamefont {Soluyanov}\ \emph {et~al.}(2015)\citenamefont
  {Soluyanov}, \citenamefont {Gresch}, \citenamefont {Wang}, \citenamefont
  {Wu}, \citenamefont {Troyer}, \citenamefont {Dai},\ and\ \citenamefont
  {Bernevig}}]{Soluyanov2015WTe2}%
  \BibitemOpen
  \bibfield  {author} {\bibinfo {author} {\bibfnamefont {Alexey~A}\
  \bibnamefont {Soluyanov}}, \bibinfo {author} {\bibfnamefont {Dominik}\
  \bibnamefont {Gresch}}, \bibinfo {author} {\bibfnamefont {Zhijun}\
  \bibnamefont {Wang}}, \bibinfo {author} {\bibfnamefont {Quansheng}\
  \bibnamefont {Wu}}, \bibinfo {author} {\bibfnamefont {Matthias}\ \bibnamefont
  {Troyer}}, \bibinfo {author} {\bibfnamefont {Xi}~\bibnamefont {Dai}}, \ and\
  \bibinfo {author} {\bibfnamefont {B.~Andrei}\ \bibnamefont {Bernevig}},\
  }\bibfield  {title} {\enquote {\bibinfo {title} {{Type-II Weyl
  semimetals}},}\ }\href@noop {} {\bibfield  {journal} {\bibinfo  {journal}
  {Nature}\ }\textbf {\bibinfo {volume} {527}},\ \bibinfo {pages} {495--498}
  (\bibinfo {year} {2015})}\BibitemShut {NoStop}%
\bibitem [{\citenamefont {Jiang}\ \emph {et~al.}(2017)\citenamefont {Jiang},
  \citenamefont {Liu}, \citenamefont {Sun}, \citenamefont {Yang}, \citenamefont
  {Rajamathi}, \citenamefont {Qi}, \citenamefont {Yang}, \citenamefont {Chen},
  \citenamefont {Peng}, \citenamefont {Hwang} \emph
  {et~al.}}]{jiang2017signature}%
  \BibitemOpen
  \bibfield  {author} {\bibinfo {author} {\bibfnamefont {Juan}\ \bibnamefont
  {Jiang}}, \bibinfo {author} {\bibfnamefont {ZK}~\bibnamefont {Liu}}, \bibinfo
  {author} {\bibfnamefont {Y}~\bibnamefont {Sun}}, \bibinfo {author}
  {\bibfnamefont {HF}~\bibnamefont {Yang}}, \bibinfo {author} {\bibfnamefont
  {CR}~\bibnamefont {Rajamathi}}, \bibinfo {author} {\bibfnamefont
  {YP}~\bibnamefont {Qi}}, \bibinfo {author} {\bibfnamefont {LX}~\bibnamefont
  {Yang}}, \bibinfo {author} {\bibfnamefont {C}~\bibnamefont {Chen}}, \bibinfo
  {author} {\bibfnamefont {H}~\bibnamefont {Peng}}, \bibinfo {author}
  {\bibfnamefont {CC}~\bibnamefont {Hwang}},  \emph {et~al.},\ }\bibfield
  {title} {\enquote {\bibinfo {title} {Signature of type-ii weyl semimetal
  phase in mote2},}\ }\href@noop {} {\bibfield  {journal} {\bibinfo  {journal}
  {Nat. Commun.}\ }\textbf {\bibinfo {volume} {8}},\ \bibinfo {pages} {13973}
  (\bibinfo {year} {2017})}\BibitemShut {NoStop}%
\bibitem [{\citenamefont {Vafek}\ and\ \citenamefont
  {Vishwanath}(2014)}]{Vafek2014}%
  \BibitemOpen
  \bibfield  {author} {\bibinfo {author} {\bibfnamefont {Oskar}\ \bibnamefont
  {Vafek}}\ and\ \bibinfo {author} {\bibfnamefont {Ashvin}\ \bibnamefont
  {Vishwanath}},\ }\bibfield  {title} {\enquote {\bibinfo {title} {{Dirac
  Fermions in Solids: From High-Tc Cuprates and Graphene to Topological
  Insulators and Weyl Semimetals}},}\ }\href@noop {} {\bibfield  {journal}
  {\bibinfo  {journal} {Annu. Rev. Condens. Matter Phys.}\ }\textbf {\bibinfo
  {volume} {5}},\ \bibinfo {pages} {83--112} (\bibinfo {year}
  {2014})}\BibitemShut {NoStop}%
\bibitem [{\citenamefont {Wang}\ \emph
  {et~al.}(2016{\natexlab{b}})\citenamefont {Wang}, \citenamefont {Vergniory},
  \citenamefont {Kushwaha}, \citenamefont {Hirschberger},\ and\ \citenamefont
  {Chulkov}}]{Wang_Heusler_2016}%
  \BibitemOpen
  \bibfield  {author} {\bibinfo {author} {\bibfnamefont {Zhijun~Wang}\
  \bibnamefont {Wang}}, \bibinfo {author} {\bibfnamefont {M.~G.}\ \bibnamefont
  {Vergniory}}, \bibinfo {author} {\bibfnamefont {S.}~\bibnamefont {Kushwaha}},
  \bibinfo {author} {\bibfnamefont {Max}\ \bibnamefont {Hirschberger}}, \ and\
  \bibinfo {author} {\bibfnamefont {E.~V.}\ \bibnamefont {Chulkov}},\
  }\bibfield  {title} {\enquote {\bibinfo {title} {Time-reversal-breaking weyl
  fermions in magnetic heusler alloys},}\ }\href@noop {} {\bibfield  {journal}
  {\bibinfo  {journal} {Phys. Rew. Lett.}\ }\textbf {\bibinfo {volume} {117}},\
  \bibinfo {pages} {236401} (\bibinfo {year} {2016}{\natexlab{b}})}\BibitemShut
  {NoStop}%
\bibitem [{\citenamefont {Armitage}\ \emph {et~al.}(2017)\citenamefont
  {Armitage}, \citenamefont {Mele},\ and\ \citenamefont
  {Vishwanath}}]{armitage2017weyl}%
  \BibitemOpen
  \bibfield  {author} {\bibinfo {author} {\bibfnamefont {NP}~\bibnamefont
  {Armitage}}, \bibinfo {author} {\bibfnamefont {EJ}~\bibnamefont {Mele}}, \
  and\ \bibinfo {author} {\bibfnamefont {Ashvin}\ \bibnamefont {Vishwanath}},\
  }\bibfield  {title} {\enquote {\bibinfo {title} {Weyl and dirac semimetals in
  three dimensional solids},}\ }\href@noop {} {\bibfield  {journal} {\bibinfo
  {journal} {arXiv preprint arXiv:1705.01111}\ } (\bibinfo {year}
  {2017})}\BibitemShut {NoStop}%
\bibitem [{\citenamefont {Xu}\ \emph {et~al.}(2015{\natexlab{b}})\citenamefont
  {Xu}, \citenamefont {Alidoust}, \citenamefont {Belopolski}, \citenamefont
  {Yuan}, \citenamefont {Bian}, \citenamefont {Chang}, \citenamefont {Zheng},
  \citenamefont {Strocov}, \citenamefont {Sanchez}, \citenamefont {Chang},
  \citenamefont {Zhang}, \citenamefont {Mou}, \citenamefont {Wu}, \citenamefont
  {Huang}, \citenamefont {Lee}, \citenamefont {Huang}, \citenamefont {Wang},
  \citenamefont {Bansil}, \citenamefont {Jeng}, \citenamefont {Neupert},
  \citenamefont {Kaminski}, \citenamefont {Lin}, \citenamefont {Jia},\ and\
  \citenamefont {Hasan}}]{Xu2015NbAs}%
  \BibitemOpen
  \bibfield  {author} {\bibinfo {author} {\bibfnamefont {Su-Yang}\ \bibnamefont
  {Xu}}, \bibinfo {author} {\bibfnamefont {Nasser}\ \bibnamefont {Alidoust}},
  \bibinfo {author} {\bibfnamefont {Ilya}\ \bibnamefont {Belopolski}}, \bibinfo
  {author} {\bibfnamefont {Zhujun}\ \bibnamefont {Yuan}}, \bibinfo {author}
  {\bibfnamefont {Guang}\ \bibnamefont {Bian}}, \bibinfo {author}
  {\bibfnamefont {Tay-Rong}\ \bibnamefont {Chang}}, \bibinfo {author}
  {\bibfnamefont {Hao}\ \bibnamefont {Zheng}}, \bibinfo {author} {\bibfnamefont
  {Vladimir~N.}\ \bibnamefont {Strocov}}, \bibinfo {author} {\bibfnamefont
  {Daniel~S.}\ \bibnamefont {Sanchez}}, \bibinfo {author} {\bibfnamefont
  {Guoqing}\ \bibnamefont {Chang}}, \bibinfo {author} {\bibfnamefont
  {Chenglong}\ \bibnamefont {Zhang}}, \bibinfo {author} {\bibfnamefont
  {Daixiang}\ \bibnamefont {Mou}}, \bibinfo {author} {\bibfnamefont {Yun}\
  \bibnamefont {Wu}}, \bibinfo {author} {\bibfnamefont {Lunan}\ \bibnamefont
  {Huang}}, \bibinfo {author} {\bibfnamefont {Chi-Cheng}\ \bibnamefont {Lee}},
  \bibinfo {author} {\bibfnamefont {Shin-Ming}\ \bibnamefont {Huang}}, \bibinfo
  {author} {\bibfnamefont {BaoKai}\ \bibnamefont {Wang}}, \bibinfo {author}
  {\bibfnamefont {Arun}\ \bibnamefont {Bansil}}, \bibinfo {author}
  {\bibfnamefont {Horng-Tay}\ \bibnamefont {Jeng}}, \bibinfo {author}
  {\bibfnamefont {Titus}\ \bibnamefont {Neupert}}, \bibinfo {author}
  {\bibfnamefont {Adam}\ \bibnamefont {Kaminski}}, \bibinfo {author}
  {\bibfnamefont {Hsin}\ \bibnamefont {Lin}}, \bibinfo {author} {\bibfnamefont
  {Shuang~Jia}\ \bibnamefont {Jia}}, \ and\ \bibinfo {author} {\bibfnamefont
  {M.~Zahid}\ \bibnamefont {Hasan}},\ }\bibfield  {title} {\enquote {\bibinfo
  {title} {{Discovery of a Weyl fermion state with Fermi arcs in niobium
  arsenide}},}\ }\href@noop {} {\bibfield  {journal} {\bibinfo  {journal} {Nat.
  Phys.}\ }\textbf {\bibinfo {volume} {11}},\ \bibinfo {pages} {748} (\bibinfo
  {year} {2015}{\natexlab{b}})}\BibitemShut {NoStop}%
\bibitem [{\citenamefont {Sun}\ \emph {et~al.}(2016)\citenamefont {Sun},
  \citenamefont {Zhang}, \citenamefont {Felser},\ and\ \citenamefont
  {Yan}}]{Sun2016}%
  \BibitemOpen
  \bibfield  {author} {\bibinfo {author} {\bibfnamefont {Yan}\ \bibnamefont
  {Sun}}, \bibinfo {author} {\bibfnamefont {Yang}\ \bibnamefont {Zhang}},
  \bibinfo {author} {\bibfnamefont {Claudia}\ \bibnamefont {Felser}}, \ and\
  \bibinfo {author} {\bibfnamefont {Binghai}\ \bibnamefont {Yan}},\ }\bibfield
  {title} {\enquote {\bibinfo {title} {{Strong intrinsic spin Hall effect in
  the TaAs family of Weyl semimetals}},}\ }\href@noop {} {\bibfield  {journal}
  {\bibinfo  {journal} {Phys. Rev. Lett.}\ }\textbf {\bibinfo {volume} {117}},\
  \bibinfo {pages} {146403} (\bibinfo {year} {2016})}\BibitemShut {NoStop}%
\bibitem [{\citenamefont {Burkov}\ and\ \citenamefont
  {Balents}(2011)}]{Burkov:2011de}%
  \BibitemOpen
  \bibfield  {author} {\bibinfo {author} {\bibfnamefont {A~A}\ \bibnamefont
  {Burkov}}\ and\ \bibinfo {author} {\bibfnamefont {Leon}\ \bibnamefont
  {Balents}},\ }\bibfield  {title} {\enquote {\bibinfo {title} {{Weyl Semimetal
  in a Topological Insulator Multilayer}},}\ }\href@noop {} {\bibfield
  {journal} {\bibinfo  {journal} {Phys. Rev. Lett.}\ }\textbf {\bibinfo
  {volume} {107}},\ \bibinfo {pages} {127205} (\bibinfo {year}
  {2011})}\BibitemShut {NoStop}%
\bibitem [{\citenamefont {Xu}\ \emph {et~al.}(2011)\citenamefont {Xu},
  \citenamefont {Weng}, \citenamefont {Wang}, \citenamefont {Dai},\ and\
  \citenamefont {Fang}}]{Xu2011}%
  \BibitemOpen
  \bibfield  {author} {\bibinfo {author} {\bibfnamefont {Gang}\ \bibnamefont
  {Xu}}, \bibinfo {author} {\bibfnamefont {Hongming}\ \bibnamefont {Weng}},
  \bibinfo {author} {\bibfnamefont {Zhijun}\ \bibnamefont {Wang}}, \bibinfo
  {author} {\bibfnamefont {Xi}~\bibnamefont {Dai}}, \ and\ \bibinfo {author}
  {\bibfnamefont {Zhong}\ \bibnamefont {Fang}},\ }\bibfield  {title} {\enquote
  {\bibinfo {title} {{Chern Semimetal and the Quantized Anomalous Hall Effect
  in HgCr$_{2}$Se$_{4}$}},}\ }\href@noop {} {\bibfield  {journal} {\bibinfo
  {journal} {{Phys. Rev. Lett.}}\ }\textbf {\bibinfo {volume} {107}},\ \bibinfo
  {pages} {186806} (\bibinfo {year} {2011})}\BibitemShut {NoStop}%
\bibitem [{\citenamefont {Zhang}\ \emph {et~al.}(2017)\citenamefont {Zhang},
  \citenamefont {Sun}, \citenamefont {Yang}, \citenamefont {?elezny},
  \citenamefont {Parkin}, \citenamefont {Felser},\ and\ \citenamefont
  {Yan}}]{Zhang_2017}%
  \BibitemOpen
  \bibfield  {author} {\bibinfo {author} {\bibfnamefont {Yang}\ \bibnamefont
  {Zhang}}, \bibinfo {author} {\bibfnamefont {Yan}\ \bibnamefont {Sun}},
  \bibinfo {author} {\bibfnamefont {Hao}\ \bibnamefont {Yang}}, \bibinfo
  {author} {\bibfnamefont {Jakub}\ \bibnamefont {?elezny}}, \bibinfo {author}
  {\bibfnamefont {Stuart P.~P.}\ \bibnamefont {Parkin}}, \bibinfo {author}
  {\bibfnamefont {Claudia}\ \bibnamefont {Felser}}, \ and\ \bibinfo {author}
  {\bibfnamefont {Binghai}\ \bibnamefont {Yan}},\ }\bibfield  {title} {\enquote
  {\bibinfo {title} {Strong anisotropic anomalous hall effect and spin hall
  effect in the chiral antiferromagnetic compounds mn 3 x (x = ge, sn, ga, ir,
  rh, and pt)},}\ }\href@noop {} {\bibfield  {journal} {\bibinfo  {journal}
  {Phys. Rew. B}\ }\textbf {\bibinfo {volume} {95}},\ \bibinfo {pages} {075128}
  (\bibinfo {year} {2017})}\BibitemShut {NoStop}%
\bibitem [{\citenamefont {de~Juan}\ \emph {et~al.}(2017)\citenamefont
  {de~Juan}, \citenamefont {Grushin}, \citenamefont {Morimoto},\ and\
  \citenamefont {Moore}}]{de2017quantized}%
  \BibitemOpen
  \bibfield  {author} {\bibinfo {author} {\bibfnamefont {Fernando}\
  \bibnamefont {de~Juan}}, \bibinfo {author} {\bibfnamefont {Adolfo~G}\
  \bibnamefont {Grushin}}, \bibinfo {author} {\bibfnamefont {Takahiro}\
  \bibnamefont {Morimoto}}, \ and\ \bibinfo {author} {\bibfnamefont {Joel~E}\
  \bibnamefont {Moore}},\ }\bibfield  {title} {\enquote {\bibinfo {title}
  {Quantized circular photogalvanic effect in weyl semimetals},}\ }\href@noop
  {} {\bibfield  {journal} {\bibinfo  {journal} {Nat. Commun.}\ }\textbf
  {\bibinfo {volume} {8}},\ \bibinfo {pages} {15995} (\bibinfo {year}
  {2017})}\BibitemShut {NoStop}%
\bibitem [{\citenamefont {Fukushima}\ \emph {et~al.}(2008)\citenamefont
  {Fukushima}, \citenamefont {Kharzeev},\ and\ \citenamefont
  {Warringa}}]{PhysRevD.78.074033}%
  \BibitemOpen
  \bibfield  {author} {\bibinfo {author} {\bibfnamefont {Kenji}\ \bibnamefont
  {Fukushima}}, \bibinfo {author} {\bibfnamefont {Dmitri~E.}\ \bibnamefont
  {Kharzeev}}, \ and\ \bibinfo {author} {\bibfnamefont {Harmen~J.}\
  \bibnamefont {Warringa}},\ }\bibfield  {title} {\enquote {\bibinfo {title}
  {Chiral magnetic effect},}\ }\href {\doibase 10.1103/PhysRevD.78.074033}
  {\bibfield  {journal} {\bibinfo  {journal} {Phys. Rev. D}\ }\textbf {\bibinfo
  {volume} {78}},\ \bibinfo {pages} {074033} (\bibinfo {year}
  {2008})}\BibitemShut {NoStop}%
\bibitem [{\citenamefont {Ba\ifmmode~\mbox{\c{s}}\else \c{s}\fi{}ar}\ \emph
  {et~al.}(2014)\citenamefont {Ba\ifmmode~\mbox{\c{s}}\else \c{s}\fi{}ar},
  \citenamefont {Kharzeev},\ and\ \citenamefont {Yee}}]{PhysRevB.89.035142}%
  \BibitemOpen
  \bibfield  {author} {\bibinfo {author} {\bibfnamefont {G\"ok\ifmmode
  \mbox{\c{c}}\else~\c{c}\fi{}e}\ \bibnamefont {Ba\ifmmode~\mbox{\c{s}}\else
  \c{s}\fi{}ar}}, \bibinfo {author} {\bibfnamefont {Dmitri~E.}\ \bibnamefont
  {Kharzeev}}, \ and\ \bibinfo {author} {\bibfnamefont {Ho-Ung}\ \bibnamefont
  {Yee}},\ }\bibfield  {title} {\enquote {\bibinfo {title} {Triangle anomaly in
  weyl semimetals},}\ }\href {\doibase 10.1103/PhysRevB.89.035142} {\bibfield
  {journal} {\bibinfo  {journal} {Phys. Rev. B}\ }\textbf {\bibinfo {volume}
  {89}},\ \bibinfo {pages} {035142} (\bibinfo {year} {2014})}\BibitemShut
  {NoStop}%
\bibitem [{\citenamefont {Goswami}\ \emph {et~al.}(2015)\citenamefont
  {Goswami}, \citenamefont {Sharma},\ and\ \citenamefont
  {Tewari}}]{PhysRevB.92.161110}%
  \BibitemOpen
  \bibfield  {author} {\bibinfo {author} {\bibfnamefont {Pallab}\ \bibnamefont
  {Goswami}}, \bibinfo {author} {\bibfnamefont {Girish}\ \bibnamefont
  {Sharma}}, \ and\ \bibinfo {author} {\bibfnamefont {Sumanta}\ \bibnamefont
  {Tewari}},\ }\bibfield  {title} {\enquote {\bibinfo {title} {Optical activity
  as a test for dynamic chiral magnetic effect of weyl semimetals},}\ }\href
  {\doibase 10.1103/PhysRevB.92.161110} {\bibfield  {journal} {\bibinfo
  {journal} {Phys. Rev. B}\ }\textbf {\bibinfo {volume} {92}},\ \bibinfo
  {pages} {161110} (\bibinfo {year} {2015})}\BibitemShut {NoStop}%
\bibitem [{\citenamefont {Chang}\ and\ \citenamefont
  {Yang}(2015)}]{PhysRevB.91.115203}%
  \BibitemOpen
  \bibfield  {author} {\bibinfo {author} {\bibfnamefont {Ming-Che}\
  \bibnamefont {Chang}}\ and\ \bibinfo {author} {\bibfnamefont {Min-Fong}\
  \bibnamefont {Yang}},\ }\bibfield  {title} {\enquote {\bibinfo {title}
  {Chiral magnetic effect in a two-band lattice model of weyl semimetal},}\
  }\href {\doibase 10.1103/PhysRevB.91.115203} {\bibfield  {journal} {\bibinfo
  {journal} {Phys. Rev. B}\ }\textbf {\bibinfo {volume} {91}},\ \bibinfo
  {pages} {115203} (\bibinfo {year} {2015})}\BibitemShut {NoStop}%
\bibitem [{\citenamefont {Chen}\ \emph {et~al.}(2013)\citenamefont {Chen},
  \citenamefont {Wu},\ and\ \citenamefont {Burkov}}]{PhysRevB.88.125105}%
  \BibitemOpen
  \bibfield  {author} {\bibinfo {author} {\bibfnamefont {Y.}~\bibnamefont
  {Chen}}, \bibinfo {author} {\bibfnamefont {Si}~\bibnamefont {Wu}}, \ and\
  \bibinfo {author} {\bibfnamefont {A.~A.}\ \bibnamefont {Burkov}},\ }\bibfield
   {title} {\enquote {\bibinfo {title} {Axion response in weyl semimetals},}\
  }\href {\doibase 10.1103/PhysRevB.88.125105} {\bibfield  {journal} {\bibinfo
  {journal} {Phys. Rev. B}\ }\textbf {\bibinfo {volume} {88}},\ \bibinfo
  {pages} {125105} (\bibinfo {year} {2013})}\BibitemShut {NoStop}%
\bibitem [{\citenamefont {Vazifeh}\ and\ \citenamefont
  {Franz}(2013)}]{PhysRevLett.111.027201}%
  \BibitemOpen
  \bibfield  {author} {\bibinfo {author} {\bibfnamefont {M.~M.}\ \bibnamefont
  {Vazifeh}}\ and\ \bibinfo {author} {\bibfnamefont {M.}~\bibnamefont
  {Franz}},\ }\bibfield  {title} {\enquote {\bibinfo {title} {Electromagnetic
  response of weyl semimetals},}\ }\href {\doibase
  10.1103/PhysRevLett.111.027201} {\bibfield  {journal} {\bibinfo  {journal}
  {Phys. Rev. Lett.}\ }\textbf {\bibinfo {volume} {111}},\ \bibinfo {pages}
  {027201} (\bibinfo {year} {2013})}\BibitemShut {NoStop}%
\bibitem [{\citenamefont {{Rees}}\ \emph {et~al.}(2019)\citenamefont {{Rees}},
  \citenamefont {{Manna}}, \citenamefont {{Lu}}, \citenamefont {{Morimoto}},
  \citenamefont {{Borrmann}}, \citenamefont {{Felser}}, \citenamefont
  {{Moore}}, \citenamefont {{Torchinsky}},\ and\ \citenamefont
  {{Orenstein}}}]{RhSiCPGE}%
  \BibitemOpen
  \bibfield  {author} {\bibinfo {author} {\bibfnamefont {Dylan}\ \bibnamefont
  {{Rees}}}, \bibinfo {author} {\bibfnamefont {Kaustuv}\ \bibnamefont
  {{Manna}}}, \bibinfo {author} {\bibfnamefont {Baozhu}\ \bibnamefont {{Lu}}},
  \bibinfo {author} {\bibfnamefont {Takahiro}\ \bibnamefont {{Morimoto}}},
  \bibinfo {author} {\bibfnamefont {Horst}\ \bibnamefont {{Borrmann}}},
  \bibinfo {author} {\bibfnamefont {Claudia}\ \bibnamefont {{Felser}}},
  \bibinfo {author} {\bibfnamefont {J.~E.}\ \bibnamefont {{Moore}}}, \bibinfo
  {author} {\bibfnamefont {Darius~H.}\ \bibnamefont {{Torchinsky}}}, \ and\
  \bibinfo {author} {\bibfnamefont {J.}~\bibnamefont {{Orenstein}}},\
  }\bibfield  {title} {\enquote {\bibinfo {title} {{Quantized Photocurrents in
  the Chiral Multifold Fermion System RhSi}},}\ }\href@noop {} {\bibfield
  {journal} {\bibinfo  {journal} {arXiv e-prints}\ ,\ \bibinfo {eid}
  {arXiv:1902.03230}} (\bibinfo {year} {2019})},\ \Eprint
  {http://arxiv.org/abs/1902.03230} {arXiv:1902.03230 [cond-mat.mes-hall]}
  \BibitemShut {NoStop}%
\bibitem [{\citenamefont {Huang}\ \emph {et~al.}(2016)\citenamefont {Huang},
  \citenamefont {Xu}, \citenamefont {Belopolski}, \citenamefont {Lee},
  \citenamefont {Chang}, \citenamefont {Chang}, \citenamefont {Wang},
  \citenamefont {Alidoust}, \citenamefont {Bian}, \citenamefont {Neupane},
  \citenamefont {Sanchez}, \citenamefont {Zheng}, \citenamefont {Jeng},
  \citenamefont {Bansil}, \citenamefont {Neupert}, \citenamefont {Lin},\ and\
  \citenamefont {Hasan}}]{Huang02022016}%
  \BibitemOpen
  \bibfield  {author} {\bibinfo {author} {\bibfnamefont {Shin-Ming}\
  \bibnamefont {Huang}}, \bibinfo {author} {\bibfnamefont {Su-Yang}\
  \bibnamefont {Xu}}, \bibinfo {author} {\bibfnamefont {Ilya}\ \bibnamefont
  {Belopolski}}, \bibinfo {author} {\bibfnamefont {Chi-Cheng}\ \bibnamefont
  {Lee}}, \bibinfo {author} {\bibfnamefont {Guoqing}\ \bibnamefont {Chang}},
  \bibinfo {author} {\bibfnamefont {Tay-Rong}\ \bibnamefont {Chang}}, \bibinfo
  {author} {\bibfnamefont {BaoKai}\ \bibnamefont {Wang}}, \bibinfo {author}
  {\bibfnamefont {Nasser}\ \bibnamefont {Alidoust}}, \bibinfo {author}
  {\bibfnamefont {Guang}\ \bibnamefont {Bian}}, \bibinfo {author}
  {\bibfnamefont {Madhab}\ \bibnamefont {Neupane}}, \bibinfo {author}
  {\bibfnamefont {Daniel}\ \bibnamefont {Sanchez}}, \bibinfo {author}
  {\bibfnamefont {Hao}\ \bibnamefont {Zheng}}, \bibinfo {author} {\bibfnamefont
  {Horng-Tay}\ \bibnamefont {Jeng}}, \bibinfo {author} {\bibfnamefont {Arun}\
  \bibnamefont {Bansil}}, \bibinfo {author} {\bibfnamefont {Titus}\
  \bibnamefont {Neupert}}, \bibinfo {author} {\bibfnamefont {Hsin}\
  \bibnamefont {Lin}}, \ and\ \bibinfo {author} {\bibfnamefont {M.~Zahid}\
  \bibnamefont {Hasan}},\ }\bibfield  {title} {\enquote {\bibinfo {title} {New
  type of weyl semimetal with quadratic double weyl fermions},}\ }\href
  {\doibase 10.1073/pnas.1514581113} {\bibfield  {journal} {\bibinfo  {journal}
  {Proc. Natl Acad. Sci. USA}\ }\textbf {\bibinfo {volume} {113}},\ \bibinfo
  {pages} {1180--1185} (\bibinfo {year} {2016})}\BibitemShut {NoStop}%
\bibitem [{\citenamefont {Chang}\ \emph {et~al.}(2016)\citenamefont {Chang},
  \citenamefont {Singh}, \citenamefont {Xu}, \citenamefont {Bian},
  \citenamefont {Huang}, \citenamefont {Hsu}, \citenamefont {Belopolski},
  \citenamefont {Alidoust}, \citenamefont {Sanchez}, \citenamefont {Zheng}
  \emph {et~al.}}]{chang2016theoretical}%
  \BibitemOpen
  \bibfield  {author} {\bibinfo {author} {\bibfnamefont {Guoqing}\ \bibnamefont
  {Chang}}, \bibinfo {author} {\bibfnamefont {Bahadur}\ \bibnamefont {Singh}},
  \bibinfo {author} {\bibfnamefont {Su-Yang}\ \bibnamefont {Xu}}, \bibinfo
  {author} {\bibfnamefont {Guang}\ \bibnamefont {Bian}}, \bibinfo {author}
  {\bibfnamefont {Shin-Ming}\ \bibnamefont {Huang}}, \bibinfo {author}
  {\bibfnamefont {Chuang-Han}\ \bibnamefont {Hsu}}, \bibinfo {author}
  {\bibfnamefont {Ilya}\ \bibnamefont {Belopolski}}, \bibinfo {author}
  {\bibfnamefont {Nasser}\ \bibnamefont {Alidoust}}, \bibinfo {author}
  {\bibfnamefont {Daniel~S}\ \bibnamefont {Sanchez}}, \bibinfo {author}
  {\bibfnamefont {Hao}\ \bibnamefont {Zheng}},  \emph {et~al.},\ }\bibfield
  {title} {\enquote {\bibinfo {title} {Theoretical prediction of magnetic and
  noncentrosymmetric weyl fermion semimetal states in the r-al-x family of
  compounds (r= rare earth, al= aluminium, x= si, ge)},}\ }\href@noop {}
  {\bibfield  {journal} {\bibinfo  {journal} {arXiv preprint arXiv:1604.02124}\
  } (\bibinfo {year} {2016})}\BibitemShut {NoStop}%
\bibitem [{\citenamefont {Bradlyn}\ \emph {et~al.}(2016)\citenamefont
  {Bradlyn}, \citenamefont {Cano}, \citenamefont {Wang}, \citenamefont
  {Vergniory}, \citenamefont {Felser}, \citenamefont {Cava},\ and\
  \citenamefont {Bernevig}}]{Bradlynaaf5037}%
  \BibitemOpen
  \bibfield  {author} {\bibinfo {author} {\bibfnamefont {Barry}\ \bibnamefont
  {Bradlyn}}, \bibinfo {author} {\bibfnamefont {Jennifer}\ \bibnamefont
  {Cano}}, \bibinfo {author} {\bibfnamefont {Zhijun}\ \bibnamefont {Wang}},
  \bibinfo {author} {\bibfnamefont {M.~G.}\ \bibnamefont {Vergniory}}, \bibinfo
  {author} {\bibfnamefont {C.}~\bibnamefont {Felser}}, \bibinfo {author}
  {\bibfnamefont {R.~J.}\ \bibnamefont {Cava}}, \ and\ \bibinfo {author}
  {\bibfnamefont {B.~Andrei}\ \bibnamefont {Bernevig}},\ }\bibfield  {title}
  {\enquote {\bibinfo {title} {Beyond dirac and weyl fermions: Unconventional
  quasiparticles in conventional crystals},}\ }\href {\doibase
  10.1126/science.aaf5037} {\bibfield  {journal} {\bibinfo  {journal}
  {Science}\ }\textbf {\bibinfo {volume} {353}} (\bibinfo {year} {2016}),\
  10.1126/science.aaf5037}\BibitemShut {NoStop}%
\bibitem [{\citenamefont {Tang}\ \emph {et~al.}(2017)\citenamefont {Tang},
  \citenamefont {Zhou},\ and\ \citenamefont {Zhang}}]{PhysRevLett.119.206402}%
  \BibitemOpen
  \bibfield  {author} {\bibinfo {author} {\bibfnamefont {Peizhe}\ \bibnamefont
  {Tang}}, \bibinfo {author} {\bibfnamefont {Quan}\ \bibnamefont {Zhou}}, \
  and\ \bibinfo {author} {\bibfnamefont {Shou-Cheng}\ \bibnamefont {Zhang}},\
  }\bibfield  {title} {\enquote {\bibinfo {title} {Multiple types of
  topological fermions in transition metal silicides},}\ }\href {\doibase
  10.1103/PhysRevLett.119.206402} {\bibfield  {journal} {\bibinfo  {journal}
  {Phys. Rev. Lett.}\ }\textbf {\bibinfo {volume} {119}},\ \bibinfo {pages}
  {206402} (\bibinfo {year} {2017})}\BibitemShut {NoStop}%
\bibitem [{\citenamefont {Pshenay-Severin}\ \emph {et~al.}(2017)\citenamefont
  {Pshenay-Severin}, \citenamefont {Ivanov}, \citenamefont {Burkov},\ and\
  \citenamefont {Burkov}}]{pshenay2017band}%
  \BibitemOpen
  \bibfield  {author} {\bibinfo {author} {\bibfnamefont {DA}~\bibnamefont
  {Pshenay-Severin}}, \bibinfo {author} {\bibfnamefont {Yu~V}\ \bibnamefont
  {Ivanov}}, \bibinfo {author} {\bibfnamefont {AA}~\bibnamefont {Burkov}}, \
  and\ \bibinfo {author} {\bibfnamefont {AT}~\bibnamefont {Burkov}},\
  }\bibfield  {title} {\enquote {\bibinfo {title} {Band structure and
  unconventional electronic topology of cosi},}\ }\href@noop {} {\bibfield
  {journal} {\bibinfo  {journal} {arXiv preprint arXiv:1710.07359}\ } (\bibinfo
  {year} {2017})}\BibitemShut {NoStop}%
\bibitem [{\citenamefont {Chang}\ \emph {et~al.}(2017)\citenamefont {Chang},
  \citenamefont {Xu}, \citenamefont {Wieder}, \citenamefont {Sanchez},
  \citenamefont {Huang}, \citenamefont {Belopolski}, \citenamefont {Chang},
  \citenamefont {Zhang}, \citenamefont {Bansil}, \citenamefont {Lin} \emph
  {et~al.}}]{chang2017large}%
  \BibitemOpen
  \bibfield  {author} {\bibinfo {author} {\bibfnamefont {Guoqing}\ \bibnamefont
  {Chang}}, \bibinfo {author} {\bibfnamefont {Su-Yang}\ \bibnamefont {Xu}},
  \bibinfo {author} {\bibfnamefont {Benjamin~J}\ \bibnamefont {Wieder}},
  \bibinfo {author} {\bibfnamefont {Daniel~S}\ \bibnamefont {Sanchez}},
  \bibinfo {author} {\bibfnamefont {Shin-Ming}\ \bibnamefont {Huang}}, \bibinfo
  {author} {\bibfnamefont {Ilya}\ \bibnamefont {Belopolski}}, \bibinfo {author}
  {\bibfnamefont {Tay-Rong}\ \bibnamefont {Chang}}, \bibinfo {author}
  {\bibfnamefont {Songtian}\ \bibnamefont {Zhang}}, \bibinfo {author}
  {\bibfnamefont {Arun}\ \bibnamefont {Bansil}}, \bibinfo {author}
  {\bibfnamefont {Hsin}\ \bibnamefont {Lin}},  \emph {et~al.},\ }\bibfield
  {title} {\enquote {\bibinfo {title} {Large fermi arcs in unconventional weyl
  semimetal rhsi},}\ }\href@noop {} {\bibfield  {journal} {\bibinfo  {journal}
  {arXiv preprint arXiv:1706.04600}\ } (\bibinfo {year} {2017})}\BibitemShut
  {NoStop}%
\bibitem [{\citenamefont {Oliveria}\ \emph {et~al.}(1988)\citenamefont
  {Oliveria}, \citenamefont {McMullan},\ and\ \citenamefont
  {Wuensch}}]{Ag2Se_cubic}%
  \BibitemOpen
  \bibfield  {author} {\bibinfo {author} {\bibfnamefont {M}~\bibnamefont
  {Oliveria}}, \bibinfo {author} {\bibfnamefont {RK}~\bibnamefont {McMullan}},
  \ and\ \bibinfo {author} {\bibfnamefont {BJ}~\bibnamefont {Wuensch}},\
  }\bibfield  {title} {\enquote {\bibinfo {title} {Single crystal neutron
  diffraction analysis of the cation distribution in the high-temperature
  phases $\alpha$-cu2-xs, $\alpha$-cu2-xse, and $\alpha$-ag2se},}\ }\href@noop
  {} {\bibfield  {journal} {\bibinfo  {journal} {Solid State Ionics}\ }\textbf
  {\bibinfo {volume} {28}},\ \bibinfo {pages} {1332--1337} (\bibinfo {year}
  {1988})}\BibitemShut {NoStop}%
\bibitem [{\citenamefont {Asadov}\ \emph {et~al.}(2015)\citenamefont {Asadov},
  \citenamefont {Aliyev},\ and\ \citenamefont {Babaev}}]{Ag2Se_review}%
  \BibitemOpen
  \bibfield  {author} {\bibinfo {author} {\bibfnamefont {Yu~G}\ \bibnamefont
  {Asadov}}, \bibinfo {author} {\bibfnamefont {Yu~I}\ \bibnamefont {Aliyev}}, \
  and\ \bibinfo {author} {\bibfnamefont {AG}~\bibnamefont {Babaev}},\
  }\bibfield  {title} {\enquote {\bibinfo {title} {Polymorphic transformations
  in cu 2 se, ag 2 se, agcuse and the role of partial cation-cation and
  anion-anion replacement in stabilizing their modifications},}\ }\href@noop {}
  {\bibfield  {journal} {\bibinfo  {journal} {Physics of Particles and Nuclei}\
  }\textbf {\bibinfo {volume} {46}},\ \bibinfo {pages} {452--474} (\bibinfo
  {year} {2015})}\BibitemShut {NoStop}%
\bibitem [{\citenamefont {Baer}\ \emph {et~al.}(1962)\citenamefont {Baer},
  \citenamefont {Busch}, \citenamefont {Fr{\"o}hlich},\ and\ \citenamefont
  {Steigmeier}}]{Ag2Se_SG17_1}%
  \BibitemOpen
  \bibfield  {author} {\bibinfo {author} {\bibfnamefont {Y}~\bibnamefont
  {Baer}}, \bibinfo {author} {\bibfnamefont {G}~\bibnamefont {Busch}}, \bibinfo
  {author} {\bibfnamefont {C}~\bibnamefont {Fr{\"o}hlich}}, \ and\ \bibinfo
  {author} {\bibfnamefont {E}~\bibnamefont {Steigmeier}},\ }\bibfield  {title}
  {\enquote {\bibinfo {title} {W{\"a}rmeleitf{\"a}higkeit, elektrische
  leitf{\"a}higkeit, hall-effekt, thermospannung und spezifische w{\"a}rme von
  ag2se},}\ }\href@noop {} {\bibfield  {journal} {\bibinfo  {journal}
  {Zeitschrift f{\"u}r Naturforschung A}\ }\textbf {\bibinfo {volume} {17}},\
  \bibinfo {pages} {886--889} (\bibinfo {year} {1962})}\BibitemShut {NoStop}%
\bibitem [{\citenamefont {Wiegers}(1971)}]{Ag2Se_SG19_1}%
  \BibitemOpen
  \bibfield  {author} {\bibinfo {author} {\bibfnamefont {GA}~\bibnamefont
  {Wiegers}},\ }\bibfield  {title} {\enquote {\bibinfo {title} {The crystal
  structure of the low-temperature form of silver selenide},}\ }\href@noop {}
  {\bibfield  {journal} {\bibinfo  {journal} {American Mineralogist: Journal of
  Earth and Planetary Materials}\ }\textbf {\bibinfo {volume} {56}},\ \bibinfo
  {pages} {1882--1888} (\bibinfo {year} {1971})}\BibitemShut {NoStop}%
\bibitem [{\citenamefont {Yu}\ and\ \citenamefont {Yun}(2011)}]{Ag2Se_SG19_2}%
  \BibitemOpen
  \bibfield  {author} {\bibinfo {author} {\bibfnamefont {Jaemin}\ \bibnamefont
  {Yu}}\ and\ \bibinfo {author} {\bibfnamefont {Hoseop}\ \bibnamefont {Yun}},\
  }\bibfield  {title} {\enquote {\bibinfo {title} {{Reinvestigation of the
  low-temperature form of Ag${\sb 2}$Se (naumannite) based on single-crystal
  data}},}\ }\href {\doibase 10.1107/S1600536811028534} {\bibfield  {journal}
  {\bibinfo  {journal} {Acta Crystallographica Section E}\ }\textbf {\bibinfo
  {volume} {67}},\ \bibinfo {pages} {i45} (\bibinfo {year} {2011})}\BibitemShut
  {NoStop}%
\bibitem [{\citenamefont {Kim}\ \emph {et~al.}(2016)\citenamefont {Kim},
  \citenamefont {Hwang}, \citenamefont {Lee}, \citenamefont {Jhi},
  \citenamefont {Lee}, \citenamefont {Park}, \citenamefont {Kim}, \citenamefont
  {Kim}, \citenamefont {Doh}, \citenamefont {Kim},\ and\ \citenamefont
  {Kim}}]{Ag2Se_SG19_topo}%
  \BibitemOpen
  \bibfield  {author} {\bibinfo {author} {\bibfnamefont {Jihwan}\ \bibnamefont
  {Kim}}, \bibinfo {author} {\bibfnamefont {Ahreum}\ \bibnamefont {Hwang}},
  \bibinfo {author} {\bibfnamefont {Sang-Hoon}\ \bibnamefont {Lee}}, \bibinfo
  {author} {\bibfnamefont {Seung-Hoon}\ \bibnamefont {Jhi}}, \bibinfo {author}
  {\bibfnamefont {Sunghun}\ \bibnamefont {Lee}}, \bibinfo {author}
  {\bibfnamefont {Yun~Chang}\ \bibnamefont {Park}}, \bibinfo {author}
  {\bibfnamefont {Si-in}\ \bibnamefont {Kim}}, \bibinfo {author} {\bibfnamefont
  {Hong-Seok}\ \bibnamefont {Kim}}, \bibinfo {author} {\bibfnamefont
  {Yong-Joo}\ \bibnamefont {Doh}}, \bibinfo {author} {\bibfnamefont {Jinhee}\
  \bibnamefont {Kim}}, \ and\ \bibinfo {author} {\bibfnamefont {Bongsoo}\
  \bibnamefont {Kim}},\ }\bibfield  {title} {\enquote {\bibinfo {title}
  {Quantum electronic transport of topological surface states in $\beta$-ag2se
  nanowire},}\ }\href {\doibase 10.1021/acsnano.5b07368} {\bibfield  {journal}
  {\bibinfo  {journal} {ACS Nano}\ }\textbf {\bibinfo {volume} {10}},\ \bibinfo
  {pages} {3936--3943} (\bibinfo {year} {2016})}\BibitemShut {NoStop}%
\bibitem [{\citenamefont {Young}\ and\ \citenamefont
  {Kane}(2015)}]{PhysRevLett.115.126803}%
  \BibitemOpen
  \bibfield  {author} {\bibinfo {author} {\bibfnamefont {Steve~M.}\
  \bibnamefont {Young}}\ and\ \bibinfo {author} {\bibfnamefont {Charles~L.}\
  \bibnamefont {Kane}},\ }\bibfield  {title} {\enquote {\bibinfo {title} {Dirac
  semimetals in two dimensions},}\ }\href {\doibase
  10.1103/PhysRevLett.115.126803} {\bibfield  {journal} {\bibinfo  {journal}
  {Phys. Rev. Lett.}\ }\textbf {\bibinfo {volume} {115}},\ \bibinfo {pages}
  {126803} (\bibinfo {year} {2015})}\BibitemShut {NoStop}%
\bibitem [{\citenamefont {Kresse}\ and\ \citenamefont {Hafner}(1993)}]{vasp}%
  \BibitemOpen
  \bibfield  {author} {\bibinfo {author} {\bibfnamefont {G.}~\bibnamefont
  {Kresse}}\ and\ \bibinfo {author} {\bibfnamefont {J.}~\bibnamefont
  {Hafner}},\ }\bibfield  {title} {\enquote {\bibinfo {title} {Ab initio
  molecular dynamics for liquid metals},}\ }\href {\doibase
  10.1103/PhysRevB.47.558} {\bibfield  {journal} {\bibinfo  {journal} {Phys.
  Rev. B}\ }\textbf {\bibinfo {volume} {47}},\ \bibinfo {pages} {558--561}
  (\bibinfo {year} {1993})}\BibitemShut {NoStop}%
\bibitem [{\citenamefont {Bl\"ochl}(1994)}]{paw1}%
  \BibitemOpen
  \bibfield  {author} {\bibinfo {author} {\bibfnamefont {P.~E.}\ \bibnamefont
  {Bl\"ochl}},\ }\bibfield  {title} {\enquote {\bibinfo {title} {Projector
  augmented-wave method},}\ }\href {\doibase 10.1103/PhysRevB.50.17953}
  {\bibfield  {journal} {\bibinfo  {journal} {Phys. Rev. B}\ }\textbf {\bibinfo
  {volume} {50}},\ \bibinfo {pages} {17953--17979} (\bibinfo {year}
  {1994})}\BibitemShut {NoStop}%
\bibitem [{\citenamefont {Kresse}\ and\ \citenamefont {Joubert}(1999)}]{paw2}%
  \BibitemOpen
  \bibfield  {author} {\bibinfo {author} {\bibfnamefont {G.}~\bibnamefont
  {Kresse}}\ and\ \bibinfo {author} {\bibfnamefont {D.}~\bibnamefont
  {Joubert}},\ }\bibfield  {title} {\enquote {\bibinfo {title} {From ultrasoft
  pseudopotentials to the projector augmented-wave method},}\ }\href {\doibase
  10.1103/PhysRevB.59.1758} {\bibfield  {journal} {\bibinfo  {journal} {Phys.
  Rev. B}\ }\textbf {\bibinfo {volume} {59}},\ \bibinfo {pages} {1758--1775}
  (\bibinfo {year} {1999})}\BibitemShut {NoStop}%
\bibitem [{\citenamefont {Perdew}\ \emph {et~al.}(1996)\citenamefont {Perdew},
  \citenamefont {Burke},\ and\ \citenamefont {Ernzerhof}}]{GGA}%
  \BibitemOpen
  \bibfield  {author} {\bibinfo {author} {\bibfnamefont {John~P.}\ \bibnamefont
  {Perdew}}, \bibinfo {author} {\bibfnamefont {Kieron}\ \bibnamefont {Burke}},
  \ and\ \bibinfo {author} {\bibfnamefont {Matthias}\ \bibnamefont
  {Ernzerhof}},\ }\bibfield  {title} {\enquote {\bibinfo {title} {Generalized
  gradient approximation made simple},}\ }\href {\doibase
  10.1103/PhysRevLett.77.3865} {\bibfield  {journal} {\bibinfo  {journal}
  {Phys. Rev. Lett.}\ }\textbf {\bibinfo {volume} {77}},\ \bibinfo {pages}
  {3865--3868} (\bibinfo {year} {1996})}\BibitemShut {NoStop}%
\bibitem [{\citenamefont {Mostofi}\ \emph {et~al.}(2008)\citenamefont
  {Mostofi}, \citenamefont {Yates}, \citenamefont {Lee}, \citenamefont {Souza},
  \citenamefont {Vanderbilt},\ and\ \citenamefont
  {Marzari}}]{mostofi2008wannier90}%
  \BibitemOpen
  \bibfield  {author} {\bibinfo {author} {\bibfnamefont {Arash~A}\ \bibnamefont
  {Mostofi}}, \bibinfo {author} {\bibfnamefont {Jonathan~R}\ \bibnamefont
  {Yates}}, \bibinfo {author} {\bibfnamefont {Young-Su}\ \bibnamefont {Lee}},
  \bibinfo {author} {\bibfnamefont {Ivo}\ \bibnamefont {Souza}}, \bibinfo
  {author} {\bibfnamefont {David}\ \bibnamefont {Vanderbilt}}, \ and\ \bibinfo
  {author} {\bibfnamefont {Nicola}\ \bibnamefont {Marzari}},\ }\bibfield
  {title} {\enquote {\bibinfo {title} {wannier90: A tool for obtaining
  maximally-localised wannier functions},}\ }\href@noop {} {\bibfield
  {journal} {\bibinfo  {journal} {Computer physics communications}\ }\textbf
  {\bibinfo {volume} {178}},\ \bibinfo {pages} {685--699} (\bibinfo {year}
  {2008})}\BibitemShut {NoStop}%
\bibitem [{\citenamefont {Mostofi}\ \emph {et~al.}(2014)\citenamefont
  {Mostofi}, \citenamefont {Yates}, \citenamefont {Pizzi}, \citenamefont {Lee},
  \citenamefont {Souza}, \citenamefont {Vanderbilt},\ and\ \citenamefont
  {Marzari}}]{MOSTOFI20142309}%
  \BibitemOpen
  \bibfield  {author} {\bibinfo {author} {\bibfnamefont {Arash~A.}\
  \bibnamefont {Mostofi}}, \bibinfo {author} {\bibfnamefont {Jonathan~R.}\
  \bibnamefont {Yates}}, \bibinfo {author} {\bibfnamefont {Giovanni}\
  \bibnamefont {Pizzi}}, \bibinfo {author} {\bibfnamefont {Young-Su}\
  \bibnamefont {Lee}}, \bibinfo {author} {\bibfnamefont {Ivo}\ \bibnamefont
  {Souza}}, \bibinfo {author} {\bibfnamefont {David}\ \bibnamefont
  {Vanderbilt}}, \ and\ \bibinfo {author} {\bibfnamefont {Nicola}\ \bibnamefont
  {Marzari}},\ }\bibfield  {title} {\enquote {\bibinfo {title} {An updated
  version of wannier90: A tool for obtaining maximally-localised wannier
  functions},}\ }\href {\doibase https://doi.org/10.1016/j.cpc.2014.05.003}
  {\bibfield  {journal} {\bibinfo  {journal} {Computer Physics Communications}\
  }\textbf {\bibinfo {volume} {185}},\ \bibinfo {pages} {2309 --2310} (\bibinfo
  {year} {2014})}\BibitemShut {NoStop}%
\bibitem [{\citenamefont {Edelstein}(1990)}]{EDELSTEIN1990233}%
  \BibitemOpen
  \bibfield  {author} {\bibinfo {author} {\bibfnamefont {V.M.}\ \bibnamefont
  {Edelstein}},\ }\bibfield  {title} {\enquote {\bibinfo {title} {Spin
  polarization of conduction electrons induced by electric current in
  two-dimensional asymmetric electron systems},}\ }\href {\doibase
  https://doi.org/10.1016/0038-1098(90)90963-C} {\bibfield  {journal} {\bibinfo
   {journal} {Solid State Communications}\ }\textbf {\bibinfo {volume} {73}},\
  \bibinfo {pages} {233 -- 235} (\bibinfo {year} {1990})}\BibitemShut {NoStop}%
\bibitem [{\citenamefont {Tao}\ and\ \citenamefont {Tsymbal}(2018)}]{Tao2018}%
  \BibitemOpen
  \bibfield  {author} {\bibinfo {author} {\bibfnamefont {L.~L.}\ \bibnamefont
  {Tao}}\ and\ \bibinfo {author} {\bibfnamefont {Evgeny~Y.}\ \bibnamefont
  {Tsymbal}},\ }\bibfield  {title} {\enquote {\bibinfo {title} {Persistent spin
  texture enforced by symmetry},}\ }\href {\doibase 10.1038/s41467-018-05137-0}
  {\bibfield  {journal} {\bibinfo  {journal} {Nature Communications}\ }\textbf
  {\bibinfo {volume} {9}},\ \bibinfo {pages} {2763} (\bibinfo {year}
  {2018})}\BibitemShut {NoStop}%
\bibitem [{\citenamefont {Kourtis}\ \emph {et~al.}(2016)\citenamefont
  {Kourtis}, \citenamefont {Li}, \citenamefont {Wang}, \citenamefont
  {Yazdani},\ and\ \citenamefont {Bernevig}}]{Kourtis2016}%
  \BibitemOpen
  \bibfield  {author} {\bibinfo {author} {\bibfnamefont {Stefanos}\
  \bibnamefont {Kourtis}}, \bibinfo {author} {\bibfnamefont {Jian}\
  \bibnamefont {Li}}, \bibinfo {author} {\bibfnamefont {Zhijun}\ \bibnamefont
  {Wang}}, \bibinfo {author} {\bibfnamefont {Ali}\ \bibnamefont {Yazdani}}, \
  and\ \bibinfo {author} {\bibfnamefont {B.~Andrei}\ \bibnamefont {Bernevig}},\
  }\bibfield  {title} {\enquote {\bibinfo {title} {{Universal signatures of
  Fermi arcs in quasiparticle interference on the surface of Weyl
  semimetals}},}\ }\href@noop {} {\bibfield  {journal} {\bibinfo  {journal}
  {Phys. Rev. B}\ }\textbf {\bibinfo {volume} {93}},\ \bibinfo {pages} {041109}
  (\bibinfo {year} {2016})}\BibitemShut {NoStop}%
\end{thebibliography}%

\end{document}